\documentclass[universe,article,accept,pdftex,moreauthors]{Definitions/mdpi} 

\usepackage{xcolor}
\usepackage{color}
\usepackage{subcaption}
\usepackage{comment} 
\firstpage{1} 
\pubvolume{1}
\issuenum{1}
\articlenumber{0}
\pubyear{2026}
\copyrightyear{2026}
\externaleditor{~} 
\datereceived{31 March 2026} 
\daterevised{6 May 2026} 
\dateaccepted{8 May 2026} 
\datepublished{12 May 2026 } 
\hreflink{https://doi.org/10.3390/universe12050142}
\pdfoutput=1 

\Title{GRMHD Simulations of Magnetized Accretion Disk/Jet: Variabilities of Black Holes and Spectral Energy Distributions in Magnetic States}

\Author{Rohan Raha 
 $^{1,}$*\orcidA{}, Banibrata Mukhopadhyay $^{1,}$*\orcidB{} and Koushik Chatterjee $^{2,}$*\orcidC{}}

\AuthorNames{Rohan Raha, Banibrata Mukhopadhyay, and Koushik Chatterjee}

\address{%
$^{1}$ \quad  Department of Physics, Indian Institute of Science, 
Bangalore 560012, India\\ 
$^{2}$ \quad Canadian Institute of Theoretical Astrophysics, University of Toronto, 60 St. George Street, \mbox{Toronto, ON M5S 3H8,} Canada}

\corres{Correspondence: raharohan@iisc.ac.in (R.R.), bm@iisc.ac.in (B.M.), kchatt@cita.utoronto.ca (K.C.)}

\abstract{We perform three-dimensional general relativistic magnetohydrodynamic 
(GRMHD) simulations of a near-maximally spinning black hole (spin parameter $a = 0.998$) with 
varying initial magnetic field geometries, systematically exploring the parameter 
space connecting magnetically arrested disk (MAD), intermediate (INT), and 
standard and normal evolution (SANE) accretion states. The magnetic 
flux threading the black hole horizon emerges as the fundamental 
state variable controlling jet efficiency, flow magnetization, and radiative 
output across all three states. We introduce complementary diagnostics—broadband spectral energy distributions spanning 
radio through hard X-ray frequencies and time-resolved X-ray light curves—that together connect simulation dynamics directly to multiwavelength observables. 
The radiative output follows a clear MAD $>$ INT $>$ SANE hierarchy in 
time-averaged luminosity, mean X-ray emission, as well as variability.  Furthermore, MAD exhibits the highest fractional variability through 
quasi-periodic magnetic flux eruption events, and INT and SANE show moderate 
variability driven by episodic reconnection and stochastic MRI turbulence, 
respectively. Scaling to GRS~1915+105, Cyg~X-1, and HLX-1, we demonstrate that 
all twelve temporal classes of GRS~1915+105 map naturally 
onto our three magnetic states, Cyg~X-1's persistent hard state is reproduced 
by a sustained INT configuration, and HLX-1's extreme luminosities arise through 
efficient Blandford--Znajek extraction in MAD states scaled to higher black hole 
mass. }


\begin{document}


\section{Introduction}

Black hole X-ray binaries (BHXRBs) exhibit remarkable spectral and temporal variability, cycling through distinct accretion states characterized by different X-ray spectral shapes, timing properties, and radio jet activity \citep{Remillard2006, belloni2000}. Understanding the physical mechanisms that drive these state transitions and govern jet launching remains one of the central challenges in high-energy astrophysics. Traditional phenomenological classifications into ``hard,'' ``soft,'' and ``intermediate'' states based on spectral hardness provide useful observational categories but offer limited physical insight into the underlying accretion physics.

Among BHXRBs, GRS 1915+105 stands out as exceptionally complex, displaying twelve distinct temporal classes with unique combinations of variability patterns, spectral properties, and outflow signatures \citep{belloni2000}. These classes range from steady hard states with persistent radio jets ($\chi$ class) to highly variable states with quasi-periodic oscillations ($\rho$ class) to thermally dominated soft states with weak or no jets ($\lambda$, $\mu$, $\gamma$ classes). The system exhibits rapid transitions between classes on timescales of seconds to minutes, demonstrating that black holes can fundamentally reorganize their accretion flow structure on dynamical timescales. Previous attempts to explain this diversity have invoked distinct physical mechanisms for different classes \citep{neilsen2011, Altamirano2011, adegoke2018}, but a unified framework capable of explaining all twelve classes through a single underlying physics remains elusive.

Complementing GRS 1915+105's remarkable variability, Cyg X-1 and HLX-1 provide crucial insights across different regimes of black hole accretion. Cyg X-1, with its precisely constrained mass of $21.2 \pm 2.2 \, M_\odot$ \citep{Miller-Jones2021}, exhibits an exceptionally stable hard state characterized by persistent radio jets with steady emission at $\sim 15$ mJy and jet power \mbox{$\sim 10^{37}$ erg s$^{-1}$ \citep{stirling2001}.} This remarkable stability, maintained over decades of observation---with a notable soft-state excursion documented in 2023 
\citep{Steiner2024} representing a rare exception---contrasts sharply with GRS 1915+105's rapid variability. However, both the sources are governed by the same underlying physics 
of magnetically regulated accretion. This suggests that different 
magnetic field configurations could perhaps produce dramatically 
different observational signatures. At the opposite end of the mass spectrum, the hyper-luminous X-ray (HLX) source, HLX-1, exhibits state transitions reminiscent of stellar-mass systems but operates at luminosities reaching $10^{40}\text{--}10^{41}$ erg s$^{-1}$ \citep{Davis2011, Servillat2011}, which challenges conventional models for stellar-mass black holes, 
but naturally fits for an intermediate-mass black hole
of $M\sim 10^4\,M_\odot$, however, accreting at the sub-Eddington rate, while maintaining hard spectral states. Understanding how the same accretion physics scales from stellar-mass binaries to intermediate-mass black holes, producing both extreme stability and extreme luminosity, requires a unified theoretical framework that has remained elusive. The concept that gradual advection of poloidal magnetic flux 
to the inner disk region drives spectral state transitions and 
relativistic ejections in GRS~1915+105 was first proposed by 
\citet{Tagger2004}, whose ``magnetic flood'' scenario attributed 
the 30-minute variability cycles and the three basic spectral 
states of \citet{belloni2000} to secular flux accumulation and 
reconnection near the central object. Our work places this concept 
on a quantitative footing through three-dimensional GRMHD 
simulations that directly measure $\phi_{\rm BH}$ and connect its 
evolution to jet power, accretion energetics, and multiwavelength 
spectral signatures across the MAD, INT, and SANE accretion 
regimes.

Theoretical work on magnetically arrested disk (MAD) states has revealed that when sufficient large-scale magnetic flux threads the black hole horizon, the resulting magnetic pressure can temporarily halt accretion \citep{narayan2003, igumenshchev2003}, creating quasi-periodic flux eruption cycles\mbox{ \citep{tchekhovskoy2011, Chatterjee:2022}}. The dimensionless magnetic flux parameter $\phi_{\rm BH}$, which measures the ratio of magnetic flux to mass accretion rate, exceeds $\sim 50$ in MAD states \citep{tchekhovskoy2011}. Conversely, standard and normal evolution (SANE) accretion maintains weak large-scale fields with $\phi_{\rm BH} < 20$ \citep{Chatterjee:2022}, dominated by magnetorotational instability (MRI)-driven turbulence \citep{balbus1991} rather than organized magnetic structures. 

General relativistic magnetohydrodynamic (GRMHD) simulations have demonstrated that MAD states enable efficient extraction of black hole rotational energy through the Blandford--Znajek mechanism \citep{Blandford1977}, producing jet efficiencies $\eta_{\rm jet} > 1$ that exceed the traditional radiative efficiency limit \citep{tchekhovskoy2011, Chatterjee:2025}. However, previous simulation studies focused primarily on either MAD or SANE extremes, without systematically exploring the intermediate regime or quantifying the distinct contributions of collimated jets versus disk/wind outflows to the total energy budget. \citet{raha2026} have studied this intermediate regime in detail and provided an analysis of all three regimes of accretion. Furthermore, direct connections between simulation-predicted energetics and observed multiwavelength luminosities remain limited, particularly for the X-ray band where disk/corona emission dominates alongside radio contributions from jet.

In this work, we perform high-resolution three-dimensional GRMHD simulations of accretion onto a near-maximally spinning black hole ($a = 0.998$) with varying initial magnetic field geometries, systematically exploring the parameter space connecting MAD and SANE regimes. We demonstrate that magnetic flux evolution through three distinct states—MAD, intermediate (INT), and SANE—apparently explains the full diversity of observed black hole phenomenology, including all twelve classes of GRS 1915+105. 

Our key innovations are threefold. First, we introduce a quantitative X-ray 
power proxy $L_{\rm X, disk}$ that measures energy flux from gas-pressure dominated 
disk and wind regions ($\sigma_m < 1$) separately from magnetically dominated 
jets ($\sigma_m > 1$), enabling direct comparison with observed X-ray ($L_X$) 
and radio ($L_R$) luminosities. Second, we post-process time-averaged GRMHD 
data to compute broadband spectral energy distributions (SEDs) spanning radio 
through hard X-ray frequencies, incorporating thermal synchrotron emission and 
synchrotron self-Compton scattering from disk and wind regions, providing 
testable spectral predictions across the black hole mass spectrum and directly 
connecting simulation dynamics to multiwavelength observables. Third, we compute 
time-resolved X-ray light curves from the spectral outputs over the quasi-steady 
interval \mbox{$t = 20{,}000$--$25{,}000\,r_g/c$,} quantifying the temporal variability 
character of each magnetic state and connecting the simulated quasi-periodic flux 
eruption events in MAD configurations to the observed variability of specific 
GRS~1915+105 classes. We scale our simulation results to GRS~1915+105, Cyg~X-1, 
and the intermediate-mass black hole candidate HLX-1, demonstrating wide 
applicability across the black hole mass spectrum. The X-ray versus radio power 
correlation emerging from our simulations reproduces observed correlations, 
providing a physical interpretation for the empirical relation 
$L_X \propto L_R^{0.5\text{--}0.7}$, in terms of combined magnetic states.

This paper is organized as follows: Section~\ref{sec:obs} discusses the observational properties of GRS~1915+105 and complementary systems Cyg~X-1 and HLX-1. Section~\ref{sec:methods} describes our numerical methods, simulation setup, and diagnostic calculations including the jet and X-ray power definitions. Section~\ref{sec:results} presents temporal evolution, flow structure, 
magnetization topology, energetics, spectral energy distribution of MAD, INT and SANE simulations, and their implications for GRS~1915+105, Cyg~X-1, and HLX-1. Section~\ref{sec:grs_all} classifies the temporal classes of GRS~1915+105 into MAD, INT and SANE. It also presents time-resolved X-ray light curves and produces a generalized correlation between X-ray and radio power across sources. Section~\ref{sec:discussion} discusses the unified physical framework, jet-disk connection, observational predictions, and limitations of our work. Section~\ref{sec:conclusions} summarizes our findings and outlines future directions. Throughout this work, we adopt geometrized units with $G = c = 1$ unless otherwise specified, and use the convention that the metric signature is $(-,+,+,+)$.

\section{Observational Properties} \label{sec:obs}

We demonstrate the temporal behaviors of the sources GRS~1915+105, Cyg X-1 and HLX-1 with our simulations. Before embarking onto simulation results, we briefly describe some important properties of these sources.

\subsection{GRS~1915+105 Temporal Classes}

GRS~1915+105 exhibits unprecedented variability divided into twelve distinct temporal classes \citep{belloni2000}, each displaying unique combinations of spectral and timing properties. Recent observations have refined its fundamental parameters: mass $12.4 \pm 2.0 \, M_\odot$, spin \mbox{$a > 0.98$ \cite{McClintock2006},} and distance $8.6 \pm 2.0$ kpc \citep{Reid2014,Miller2020}, making it an ideal laboratory for studying extreme accretion physics. The source exhibits unusually bright and steady emission independent of its spectral nature. We suggest, as shown below by our simulations, that this persistent luminosity is due to its magnetically driven power extraction.

Its $\chi$ class merits particular attention due to its complex substructure. While initially appearing as a steady hard state, detailed analysis by \citet{belloni2000} revealed four distinct subclasses ($\chi_1$, $\chi_2$, $\chi_3$, $\chi_4$). The $\chi_1$ and $\chi_3$ classes can be modeled with a simple power-law spectrum: spectral index (SI)$\sim 3.0$, while $\chi_2$ and $\chi_4$ show systematically lower absorption and harder spectra \citep{belloni2000}.  The steady radio emission in $\chi$ classes \citep{Klein-Wolt2002} correlates with the strongest magnetic field configurations in our MAD simulations, explaining both the spectral hardness and jet stability.

The $\rho$ class exhibits dramatic ``heartbeat'' oscillations \citep{neilsen2011}, characterized by disk inflation followed by rapid collapse and discrete ejection events. \citet{adegoke2018} showed that this temporal class maintains significant sub-Keplerian/coronal contribution ($72\%$ power-law, $28\%$ diskbb---multi-temperature 
blackbody emission from a geometrically thin Keplerian 
disk) despite its oscillatory behavior. \citet{Neilsen2012} demonstrated that these cycles involve complex interplay in magnetic processes, which our simulations reproduce through periodic magnetic flux accumulation and eruption.

The $\alpha$ class stands out for exhibiting the hardest spectrum among all classes with SI $\sim 2.39$ and the lowest luminosity ($0.1341 \, L_{\rm Edd}$) \citep{adegoke2018}. Its strongly power-law dominated spectrum ($77\%$) with minimal diskbb contribution ($23\%$)  and fractal nature of lightcurve\mbox{ \citep{adegoke2018}} represents the closest approximation to a pure advection dominated accretion flow (ADAF) state \citep{narayan1994}. Our simulations show that this configuration arises in low-accretion rate regimes where magnetic fields create a hot, optically thin flow with efficient angular momentum\mbox{ transport.}

The $\theta$ class presents a particularly interesting case that has long challenged theoretical models. Despite switching between hard and soft states with high power-law dominance ($88\%$), which suggests a hard state, it maintains quite high luminosity ($0.3536 \, L_{\rm Edd}$) \citep{adegoke2018}. \citet{belloni2000} identified it as a peculiar combination of spectral states A and C with distinctive high-frequency quasi-periodic oscillations (QPOs). Our INT simulations demonstrate how this unusual combination can arise when intermediate magnetic fields maintain efficient disk accretion while driving substantial outflows, explaining both the hard spectrum and high luminosity.

The $\beta$ class displays the most rapid spectral transitions, with nearly equal diskbb and power-law contributions ($46\%$ diskbb, $52\%$ power-law, \citep{adegoke2018}) and steep SI $\sim 3.25$ \citep{Migliari2003}. These transitions occur on timescales of seconds \citep{belloni2000}, challenging traditional viscous timescale arguments. Our MAD simulations demonstrate how rapid magnetic field reconfiguration can drive such fast transitions, providing an explanation for this puzzling behavior.

Particularly intriguing are the intermediate classes $\kappa$ and $\nu$, which often precede major transitions \citep{Klein-Wolt2002}. The $\kappa$ class shows balanced diskbb ($49\%$) and power-law ($51\%$) contributions with SI $\sim 2.99$, while the $\nu$ class exhibits stronger power-law dominance ($72\%$) with SI $\sim 2.93$ \citep{adegoke2018}. Timing analysis by \citet{belloni2013} 
revealed that these states mark transitions in the underlying magnetic field structure, consistent with our simulations of INT state between MAD and SANE configurations.

The $\mu$ and $\lambda$ classes present an interesting case of diskbb dominance (>54\%) combined with strong variability and steep spectrum (SI $>2.9$) \citep{adegoke2018}. The $\gamma$, $\phi$, and $\delta$ classes exhibit stochastic variability \citep{adegoke2018} with high diskbb contributions and steep spectra (SI $> 3.4$), argued to be radiation-trapped accretion flows. The $\gamma$ class shows particularly high diskbb contribution ($60\%$) with stochastic behaviour in time series \citep{adegoke2018}, suggesting them to corroborate with our simulated SANE configuration, particularly in terms of its temporal variability.

\subsection{Multi-Wavelength Properties of GRS~1915+105}

High-resolution X-ray spectroscopy has revealed sophisticated wind structures with velocities reaching $\sim 1000$ km/s \citep{Neilsen2012}.
These winds show remarkable anti-correlation with jet activity \citep{Miller2016}. In our simulations, during MAD states, vertical fields collimate jets, while in SANE states, magnetic turbulence drives wider-angle winds.

Radio observations reveal two distinct types of jets: steady, compact jets in hard states with flat spectra, and discrete ejection events during state transitions with steep spectra \citep{fender2004}. The jet power varies from $10^{38}$ to $10^{39}$ erg/s, with higher values during major ejection events. Our simulations demonstrate different accretion states connecting above mentioned jet modes revealed by varying magnetic field configurations. However, the current nonradiative, hence
idealistic, simulations do not seem to distinguish GRS 1915+105 
classes with steady jets, e.g., $\chi$ \citep{Rushton2010} from the ones with episodic jets, e.g., $\kappa$, $\theta$, $\beta$, etc. \citep{Klein-Wolt2002, Neilsen2010}. 
However, classes $\gamma,~\delta,~\phi$ and $~\mu$ exhibiting weak jets \citep{Klein-Wolt2002} are identified as SANE state. The observed jets are very highly correlated with hard X-ray emission and are often characterized by a peak in radio flux which generally follows a dip in X-ray emission \citep{Mirabel1998}.

Moreover, QPOs show systematic evolution with spectral state. Low-frequency (\mbox{0.1--10 Hz}) QPOs dominate hard states \citep{Rodriguez2008}, while high-frequency (\mbox{100--450 Hz}) QPOs appear in intermediate states \citep{belloni2000}. 

\subsection{Complementary Systems}

Cyg X-1, with its precisely measured mass of $21.2 \pm 2.2 \, M_\odot$ \citep{Miller-Jones2021}, provides crucial comparative insights. Its persistent hard-state jet shows remarkable stability, with steady radio emission at $\sim $15 mJy and jet power $\sim 10^{37}$ erg/s \citep{stirling2001}. Our simulations suggest this stability arises from maintaining an intermediate magnetic field strength, avoiding the extreme variability of MAD states while still supporting jet production.

Further, the HLX source HLX-1 extends our understanding to higher mass scales ($\sim 10^4 \, M_\odot$). Its state transitions mirror those seen in GRS~1915+105 but on longer timescales\mbox{ \citep{Servillat2011}}. HLX-1's achieving hard state luminosity of $10^{40}$--$10^{41}$ erg/s  at sub-Eddington accretion rates \citep{Davis2011} challenges conventional models; however, this is naturally explained by our MAD simulations through efficient extraction of black hole spin energy
.

The timing--spectral--jet connections observed across this mass range ($10$--$10^5 \, M_\odot$) suggest common underlying physics scaling with black hole mass. Our simulations of different magnetic field configurations provide a unified framework explaining the various observational challenges: extreme variability, stable jets, and super-Eddington luminosities.

\section{Methods}\label{sec:methods}

\subsection{Numerical Setup and Initial Conditions}

We perform high-resolution three-dimensional (3D) GRMHD simulations using the H-AMR code \citep{liska2022} to investigate accretion flows around a rapidly spinning black hole with dimensionless spin parameter $a = 0.998$. This near-maximal spin is physically motivated by 
observational constraints: GRS~1915+105 has a measured 
spin $a > 0.98$ \citep{McClintock2006}, and Cyg~X-1 is 
similarly constrained to be near-maximally spinning 
($a \gtrsim 0.98$ \citep{Gou2011, Fabian2012}). 
For HLX-1 the spin remains observationally unconstrained, 
but near-maximal spin is required to explain its extreme 
luminosities through efficient Blandford--Znajek extraction 
at sub-Eddington accretion rates. The qualitative 
MAD/INT/SANE state classification and the role of 
$\phi_{\rm BH}$ as the fundamental state variable are 
robust to spin; spin primarily sets the scale of 
$\eta_{\rm jet}$ rather than determining the state 
identity. A systematic survey of spin dependence is 
reserved for future work (see Section~\ref{sec:discussion}). The code solves the ideal GRMHD equations in Kerr spacetime using a conservative finite-volume scheme that maintains both the divergence-free constraint on magnetic fields and conservation of mass-energy-momentum to machine\mbox{ precision.}

The mass continuity equation is:
\begin{equation}
\nabla_\mu (\rho u^\mu) = 0,
\end{equation}
\noindent where $\rho$ is the rest-mass density and $u^\mu$ is the four-velocity. Energy-momentum conservation follows:
\begin{equation}
\nabla_\mu T^{\mu\nu} = 0,
\end{equation}
\noindent with the stress-energy tensor in ideal GRMHD is:
\begin{equation}
T^{\mu\nu} = (\rho + u_{\rm gas} + p_{\rm gas} + b^2) u^\mu u^\nu + \left(p_{\rm gas} + \frac{b^2}{2}\right)g^{\mu\nu} - b^\mu b^\nu.
\label{eq:tmunu}
\end{equation}
\noindent This
can be decomposed into hydrodynamic and electromagnetic components, respectively:
\begin{align}
T^{\mu\nu}_{\rm HD} &= (\rho + u_{\rm gas} + p_{\rm gas}) u^\mu u^\nu + p_{\rm gas} g^{\mu\nu}, \label{eq:tmunu_hd} \\
T^{\mu\nu}_{\rm EM} &= b^2 u^\mu u^\nu + \frac{b^2}{2} g^{\mu\nu} - b^\mu b^\nu. \label{eq:tmunu_em}
\end{align}
\noindent where $u_{\rm gas}$ is the internal energy density, $p_{\rm gas}$ is the gas pressure, $b^\mu$ is the magnetic field four-vector with $b^2 = b_\mu b^\mu$, and $g^{\mu\nu}$ is the contravariant metric tensor. The magnetic field evolves through the ideal MHD induction equation: 
\begin{equation}
\nabla_\mu (^* F^{\mu\nu}) = 0,
\end{equation}
\noindent where $^*F^{\mu\nu}$ is the dual of the electromagnetic field tensor, ensuring the magnetic field remains frozen into the fluid. The divergence-free constraint $\nabla\cdot B=0$, is maintained via constrained transport, where $B$ is the 3-magnetic field defined as 
\begin{align}
b^t = B^i u^\mu g_{i\mu},\\ \nonumber
b^i = \frac{B^i + b^t u^i}{u^t}
\end{align}
\noindent We close the system with an ideal gas equation of state $p_{\rm gas} = (\Gamma - 1) u_{\rm gas}$, where $\Gamma$ is the adiabatic index appropriate for a relativistic plasma.

The computational domain extends from inner radius, $r_{\rm in} = 0.87r_H$, to outer radius, $r_{\rm out} = 10^6 r_g$, where $r_H = r_g(1 + \sqrt{1-a^2})$ is the event horizon radius and $r_g = GM_{\rm BH}/c^2$ is the gravitational radius. The choice $r_{\rm in} = 0.87\,r_H$ ensures 
five grid cells inside the event horizon, as required 
by the piecewise parabolic method (PPM) interpolation 
scheme employed in H-AMR \citep{liska2022}. For our 3D simulations, we employ a grid resolution of \mbox{$N_r \times N_\theta \times N_\phi = 448 \times 240 \times 192$.} This resolution is chosen to adequately resolve both the magnetorotational instability (MRI) that drives turbulent transport and the large-scale magnetic structures that develop in MAD states. We apply outflowing radial boundary conditions, transmissive polar boundary conditions, and periodic boundary conditions in the azimuthal direction.

Each simulation is initialized with a Fishbone--Moncrief equilibrium torus \citep{Fishbone1976} with inner edge at $r_{\rm in} = 20r_g$ and pressure maximum at $r_{\rm max} = 41r_g$. The pressure maximum at $r_{\rm max} = 41\,r_g$ 
follows standard community practice 
\citep{Fishbone1976, McKinney2012} and corresponds 
to the radius of maximum gas pressure in the 
hydrostatic Fishbone--Moncrief solution. This choice 
yields a torus extending to $\sim 10^3\,r_g$, 
providing a sufficient gas reservoir for stable 
quasi-steady accretion while remaining well resolved 
by our computational grid. We use an ideal gas equation of state with adiabatic index $\Gamma = 13/9$, appropriate for a relativistic plasma. The initial torus is threaded with a large-scale poloidal magnetic field whose geometry and strength are varied to explore different evolutionary pathways toward distinct accretion\mbox{ states.}

To systematically explore the parameter space of magnetic field configurations, we initialize three distinct poloidal field geometries characterized by different vector potential\mbox{ distributions:}

\begin{itemize}
\item \textbf{Standard and Normal Evolution (SANE):}
 Simple poloidal field concentrated in high-density regions, with the azimuthal component of magnetic vector potential
\begin{equation}
A_\phi = \max(q, 0), \quad q = \frac{\rho}{\rho_{\rm max}} - 0.2.
\end{equation}
This configuration creates magnetic field lines confined to the densest parts of the accretion flow, concentrating flux in regions where it can be efficiently advected\mbox{ inward.}

\item \textbf{Magnetically Arrested Disk (MAD):}  Complex configuration with enhanced equatorial field strength, with
\begin{equation}
A_\phi = \max(q, 0), \quad q = \frac{\rho}{\rho_{\rm max}} \left(\frac{r}{r_{\rm in}}\right)^3 \sin^3\theta \, e^{-r/400} - 0.2.
\end{equation}
The additional radial and angular dependencies produce a stronger field near the equatorial plane that falls off with radius, promoting vertical field structures and rapid magnetic flux accumulation near the black hole.

\item \textbf{Intermediate (INT):}  Field scaling with both density and radius, with
\begin{equation}
A_\phi = \max(q^2 r^3, 0), \quad q = \frac{\rho}{\rho_{\rm max}} - 0.2.
\end{equation}
The quadratic dependence on $q$ produces stronger contrast between high and low density regions, while the $r^3$ factor drives strong magnetic activity in the outer disk regions \citep{raha2026}.
\end{itemize}

Physically, these three configurations produce 
distinct initial field topologies. The SANE setup 
creates a simple poloidal field confined to high-density 
equatorial regions with relatively weak and disordered 
field lines. The MAD configuration produces a stronger 
field concentrated near the equatorial plane with 
significant vertical structure, promoting rapid 
large-scale flux accumulation near the black hole. 
The INT configuration generates a field that scales 
with both density and radius, producing intermediate 
flux accumulation and a mixed disk--jet morphology. 
These initial geometries and their corresponding 
density distributions are illustrated in Figure~2 of 
\citet{raha2026}. As the simulations evolve, matter 
accretes inward through MRI-driven turbulent angular 
momentum transport in the disk, while magnetically 
dominated polar funnels develop self-consistently 
into collimated jets; the corona-like hot, optically 
thin gas occupying the wind region ($\sigma_m < 1$, 
$|\pi/2 - \theta| > h/r$) also develops 
self-consistently without prescription.

For each magnetic field geometry, we vary the initial plasma-$\beta = p_{\rm gas,max}/p_{\rm mag,max}$ from 0.1 to 1000, where $p_{\rm gas,max}$ and $p_{\rm mag,max}$ are the maximum gas and magnetic pressures, respectively, in the initial configuration. This range spans from strongly magnetized initial conditions (initial plasma-$\beta = 0.1$) that can rapidly form MAD states to weakly magnetized configurations (initial plasma-$\beta = 1000$) that evolve as SANE states. The simulations are labeled according to their field geometry and initial plasma-$\beta$ value (e.g., P2B100 denotes P2 geometry with initial plasma-$\beta = 100$). The full parameter survey comprises three magnetic 
field geometry families (P1, P2, and PL) each explored 
with five initial plasma-$\beta$ values ($\beta_0 = 0.1, 
1, 10, 100, 1000$), yielding 15 low-resolution 2D 
simulations whose accretion states are catalogued in 
Table~C1 and C2 of \citet{raha2026}. From this survey, 
P2B100, PLB100, and P1B100 are selected as 
representative configurations for high-resolution 3D 
analysis based on their robustly distinct 2D behaviors: 
P2B100 consistently reaches MAD saturation ($\phi_{\rm BH} 
\gtrsim 50$), PLB100 maintains intermediate magnetic 
flux ($20 \lesssim \phi_{\rm BH} \lesssim 40$), and 
P1B100 remains in a SANE state ($\phi_{\rm BH} < 20$) 
throughout.

All simulations are evolved until reaching quasi-steady state by $t \gtrsim 20{,}000 r_g/c$. We verify that steady-state conditions are achieved by examining the constancy of the mass accretion rate profile with radius, confirming inflow equilibrium out to $r \approx 60r_g$. Analysis is performed over the time interval $t = 20{,}000\text{--}25{,}000 r_g/c$, during which the flow has settled into its characteristic state (MAD, INT, or SANE) as determined by the time-averaged magnetic flux and variability properties.

\subsection{MRI Resolution}

To ensure our simulations adequately resolve the MRI that drives turbulent angular momentum transport, we calculate the quality factor $Q_{\rm MRI}$ following the methodology of \citet{Hawley2011} and \citet{McKinney2012}. The quality factor represents the number of grid cells per fastest-growing MRI wavelength in each coordinate direction:
\begin{equation}
Q_{x,\rm MRI} \equiv \frac{\lambda_{x,\rm MRI}}{\Delta x},
\label{eq:qmri}
\end{equation}
\noindent where $x \in \{r, \theta, \phi\}$, $\Delta x$ is the grid spacing in direction $x$, and the MRI wavelength is given\mbox{ by:}
\begin{equation}
\lambda_{x,\rm MRI} \approx \frac{2\pi |v_{x,A}|}{|\Omega_{\rm rot}|},
\label{eq:lambda_mri}
\end{equation}
\noindent where $v_{x,A} = \sqrt{b_x b^x/\epsilon}$ is the Alfv\'en speed in direction $x$ with $\epsilon \equiv b^2 + \rho + u_{\rm gas} + p_{\rm gas}$, and $\Omega_{\rm rot} = v_\phi/r$ is the angular velocity \citep{Chatterjee:2019}. 

Proper resolution of MRI-driven turbulence requires $Q_{\rm MRI} \gtrsim 6-10$ in all \linebreak  directions\mbox{ \citep{Sano2004,Porth:2019}}. Our simulations significantly exceed this threshold throughout the disk region ($r = 20\text{--}60 r_g$). The MAD configuration achieves exceptional resolution with $Q_{r,\rm MRI} \approx 82$, $Q_{\theta,\rm MRI} \approx 94$, and $Q_{\phi,\rm MRI} \approx 75$, reflecting the strong magnetic fields characteristic of this state. The INT and SANE configurations maintain $Q_{\rm MRI} > 10$ in all directions (INT: $Q_{r,\rm MRI} \approx 14$, $Q_{\theta,\rm MRI} \approx 14$, $Q_{\phi,\rm MRI} \approx 25$; SANE: $Q_{r,\rm MRI} \approx 11$, $Q_{\theta,\rm MRI} \approx 11$, $Q_{\phi,\rm MRI} \approx 21$), well above the minimum required for capturing MRI turbulence. The higher azimuthal resolution in all states ensures adequate sampling of non-axisymmetric modes that can contribute to angular momentum transport.

\subsection{Physical Quantities and Diagnostics}
\label{sec:diagnostics}

We compute several key physical quantities to characterize the accretion flow properties, magnetic field evolution, and outflow energetics. All quantities are calculated from simulation outputs produced every $\sim 10 r_g/c$. Time-averaged quantities $\langle Q \rangle$ are computed over the interval $t = 20{,}000\text{--}25{,}000 r_g/c$ for 3D simulations. Density-weighted averages over angular coordinates are denoted $\langle Q \rangle_\rho$, calculated as:
\begin{equation}
\langle Q \rangle_\rho(r) = \frac{\int \rho Q \sqrt{-g} \, d\theta d\phi}{\int \rho \sqrt{-g} \, d\theta d\phi},
\label{eq:density_weighted}
\end{equation}
\noindent where $g$ is the determinant of the metric tensor.

\subsubsection{Mass Accretion Rate}

The mass accretion rate as a function of radius is calculated by integrating the radial mass flux over a spherical shell:
\begin{equation}
\dot{M}(r) = -\int \rho u^r \sqrt{-g} \, d\theta d\phi,
\label{eq:mdot}
\end{equation}
\noindent where the negative sign ensures $\dot{M} > 0$ for inflow ($u^r < 0$). We normalize the code accretion rate by the Eddington value:
\begin{equation}
\dot{M}_{\rm Edd} = \frac{L_{\rm Edd}}{\eta_{\rm rad} c^2} = \frac{4\pi G M_{\rm BH} m_p}{\eta_{\rm rad} \sigma_T c} \approx 1.4 \times 10^{18} \left(\frac{M_{\rm BH}}{M_\odot}\right) \text{ g s}^{-1},
\label{eq:mdot_edd}
\end{equation}
\noindent where $L_{\rm Edd} = 4\pi G M_{\rm BH} m_p c/\sigma_T \approx 1.3 \times 10^{38} (M_{\rm BH}/M_\odot)$ erg s$^{-1}$ is the Eddington luminosity, $\eta_{\rm rad} = 0.1$ is the standard radiative efficiency, $G$ is the gravitation constant, $m_p$ is the proton mass, and $\sigma_T$ is the Thomson scattering cross section. The dimensionless accretion rate is $\dot{m} = \dot{M}/\dot{M}_{\rm Edd}$.

\subsubsection{Magnetic Flux}

The dimensionless magnetic flux threading the black hole horizon provides a crucial diagnostic for distinguishing between different accretion states \citep{tchekhovskoy2011, narayan2003}:
\begin{equation}
\phi_{\rm BH} = \frac{1}{2}\sqrt{\frac{4\pi}{\dot{M}}} \int |B^r| \sqrt{-g} \, d\theta d\phi,
\label{eq:phi}
\end{equation}
\noindent where $B^r$ is the radial component of the magnetic field measured at the event horizon $r = r_H$, and the normalization by $\sqrt{\dot{M}}$ renders $\phi_{\rm BH}$ dimensionless. For a rapidly spinning black hole, the MAD saturation threshold is approximately \citep{tchekhovskoy2011}:
\begin{equation}
\phi_{\rm MAD}(a) \approx 52.6 + 34a - 14.9a^2 - 20.2a^3.
\label{eq:phi_mad}
\end{equation}

\subsubsection{Stress-Energy Tensor and Energy Fluxes}

The radial energy flux is obtained from the $T^r_{\phantom{r}t}$ component:
\begin{equation}
\dot{E}(r) = -\int \sqrt{-g} T^r_{\phantom{r}t} \, d\theta d\phi,
\label{eq:edot}
\end{equation}
\noindent where positive values of $-T^r_{\phantom{r}t}$ indicate outward energy transport, while negative values represent inward advection. The energy flux separates into hydrodynamic and electromagnetic contributions:
\begin{align}
\dot{E}_{\rm HD}(r) &= -\int \sqrt{-g} T^r_{\phantom{r}t,\rm HD} \, d\theta d\phi, \label{eq:edot_hd} \\
\dot{E}_{\rm EM}(r) &= -\int \sqrt{-g} T^r_{\phantom{r}t,\rm EM} \, d\theta d\phi. \label{eq:edot_em}
\end{align}

\subsubsection{Disk Scale Height}
\label{sec:scale_height}

The density-weighted disk scale height-to-radius ratio is:
\begin{equation}
\left\langle \frac{h}{r} \right\rangle_\rho = \frac{\int \rho |\pi/2 - \theta| \sqrt{-g} \, d\theta d\phi}{\int \rho \sqrt{-g} \, d\theta d\phi},
\label{eq:h_over_r}
\end{equation}
\noindent where $|\pi/2 - \theta|$ measures angular distance from the equatorial plane. This provides a measure of the vertical extent of the accretion flow.

\subsubsection{Plasma-$\beta$ and Magnetization}

The plasma-$\beta$ parameter measures the ratio of gas to magnetic pressure:
\begin{equation}
\beta = \frac{p_{\rm gas}}{p_{\rm mag}} = \frac{2p_{\rm gas}}{b^2},
\label{eq:beta}
\end{equation}
\noindent where $p_{\rm mag} = b^2/2$ is the magnetic pressure. This characterizes the relative importance of thermal versus magnetic support.

The magnetization parameter is:
\begin{equation}
\sigma_m = \frac{b^2}{\rho},
\label{eq:sigma}
\end{equation}
\noindent which measures the ratio of electromagnetic to rest-mass energy density. Regions with $\sigma_m > 1$ are magnetically dominated and typically associated with jets, while $\sigma_m < 1$ indicates gas-pressure dominated disk and wind regions. Combined with disk scale height, we identify three flow regions:
\begin{itemize}
\item \textbf{Jet region:}  $\sigma_m > 1$
\item \textbf{Disk region:}  $\sigma_m < 1$ and $|\pi/2 - \theta| < h/r$
\item \textbf{Wind region:}  $\sigma_m < 1$ and $|\pi/2 - \theta| > h/r$
\end{itemize}

\subsubsection{Jet Power and Outflow Efficiency}

The efficiency of converting accreting rest-mass energy into outgoing energy flux is defined as:
\begin{equation}
\eta_{\rm jet} = \frac{\dot{E} + \dot{M}c^2}{\dot{M}c^2} = \frac{\dot{E}}{\dot{M}c^2} + 1,
\label{eq:eta_jet}
\end{equation}
\noindent where both $\dot{E}$ and $\dot{M}$ are evaluated from Equations (\ref{eq:mdot}) and (\ref{eq:edot}), respectively. This efficiency can exceed unity when black hole rotational energy is extracted via the Blandford--Znajek mechanism \citep{Blandford1977}. The physical jet power in cgs units is:
\begin{equation}
P_{\rm jet} = \eta_{\rm jet} \dot{m} \dot{M}_{\rm Edd} c^2 \approx 1.1 \times 10^{39} \, \eta_{\rm jet} \, \dot{m} \left(\frac{M_{\rm BH}}{M_\odot}\right) \text{ erg s}^{-1}.
\label{eq:pjet}
\end{equation}

\subsubsection{Spectral Energy Distribution}
\label{sec:sed_calc}

We provide testable spectral predictions by post-processing time-averaged GRMHD data to compute broadband spectral energy distributions (SEDs) from disk and wind regions ($\sigma_m < 1$). 

\textbf{Electron Thermodynamics:}
In the lieu of two-temperature plasma model, for the present purpose, we hypothesize the electron temperature $T_e = T_p/R$, where $R$ represents the proton-to-electron temperature ratio. We choose $R$ based on theoretical knowledge\mbox{ \cite{Ressler2015, Chael2017},} and it is in accordance with particle-in-cell simulations of collisionless accretion flows \citep{Howes2010}. The proton temperature $T_p$ follows from the ideal gas law:
\begin{equation}
T_p = \frac{\mu_i p_{\rm gas} m_p}{\rho k_B},
\label{eq:tp}
\end{equation}
\noindent where $\mu_i = 0.5$ is the mean molecular weight per ion for fully ionized hydrogen. The dimensionless electron temperature is $\Theta_e = k_B T_e/m_e c^2$, and the electron number density is $n_e = \rho/\mu_e m_p$ with $\mu_e = 0.5$.

\textbf{Synchrotron Emission:}
We calculate thermal synchrotron emission from relativistic Maxwellian electrons following the analytic fitting formulae of \citet{Leung2011}. The characteristic synchrotron frequency is $\nu_s = eB/(2\pi m_e c)$, where $B = \sqrt{b^2}$, is the magnetic field strength, and the critical frequency is $\nu_c = (3/2)\nu_s \Theta_e^2$. The synchrotron emissivity per unit volume is:
\begin{equation}
j_\nu^{\rm syn} = \frac{\sqrt{2}\pi e^2 n_e \nu_s}{3c K_2(1/\Theta_e)} \left[X^{1/2} + 2^{11/12}X^{1/6}\right]^2 \exp(-X^{1/3}),
\label{eq:j_syn}
\end{equation}
\noindent where $X = \nu/\nu_c$, is the dimensionless frequency and $K_2$ is the modified Bessel function of the second kind.

Synchrotron self-absorption becomes important at low frequencies and is included via the optical depth $\tau_{\rm syn} = \alpha_\nu \ell$, where $\alpha_\nu = j_\nu c^2/(2\nu^2 k_B T_e)$ is the absorption coefficient and $\ell$ is the local length scale. We calculate $\ell$ as the vertical pressure scale height:
\begin{equation}
\ell = \frac{p_{\rm gas} \, r}{|\partial p_{\rm gas}/\partial \theta|},
\label{eq:length_scale}
\end{equation}
\noindent where the gradient is computed along the polar direction. The self-absorbed emissivity is:
\begin{equation}
j_\nu^{\rm eff} = j_\nu^{\rm syn} \frac{1 - \exp(-\tau_{\rm syn})}{\tau_{\rm syn}},
\label{eq:j_eff}
\end{equation}
\noindent which smoothly interpolates between the optically thin limit ($\tau_{\rm syn} \ll 1$, $j_\nu^{\rm eff} \approx j_\nu^{\rm syn}$) and the optically thick Rayleigh--Jeans limit ($\tau_{\rm syn} \gg 1$, $j_\nu^{\rm eff} \propto B_\nu$).

\textbf{Inverse-Compton Scattering:} 
We compute synchrotron self-Compton (SSC) emission where hot electrons upscatter synchrotron seed photons through multiple inverse-Compton scatterings. The optical depth for Thomson scattering is $\tau = n_e \sigma_T \ell$, where $\sigma_T$ is the Thomson cross section. The Compton y-parameter quantifying energy transfer to photons\mbox{ is:}
\begin{equation}
y = 4\Theta_e \, \tau_{\rm eff},
\label{eq:y_compton}
\end{equation}
\noindent where $\tau_{\rm eff} = \min[\max(\tau, \tau^2), 10]$ captures both optically thin ($\tau < 1$) and thick ($\tau > 1$) regimes while preventing unphysical divergence \citep{RybickiLightman1979}.

Each scattering boosts photon energy by an average amplification factor $A = 1 + 4\Theta_e + 16\Theta_e^2$ \citep{RybickiLightman1979}. After $N \approx \tau_{\rm eff}$ scatterings (capped at physically motivated limits depending on optical depth), an observed photon at frequency $\nu_{\rm obs}$ is originated as a seed photon at $\nu_{\rm seed} = \nu_{\rm obs}/A^N$. The inverse-Compton emissivity is:
\begin{equation}
j_\nu^{\rm IC}(\nu_{\rm obs}) = y \, j_\nu^{\rm eff}(\nu_{\rm seed}) \times \exp\left(-\frac{h\nu_{\rm obs}}{f_{\rm cutoff} k_B T_e}\right),
\label{eq:j_ic}
\end{equation}
\noindent where the exponential Wien cutoff with $f_{\rm cutoff} = 2$ prevents unphysical emission at $h\nu \gg k_B T_e$ \citep{Dermer2009}.

\textbf{Luminosity Calculation:} 
Emissivities are calculated only in gas-pressure dominated disk and wind regions ($\sigma_m < 1$). This ensures direct correspondence between radiative luminosities and kinematic energy fluxes from disk and wind regions, excluding magnetically dominated jets. Spectral luminosities are computed by integrating over the computational domain:
\begin{equation}
L_\nu = \int_{\sigma_m < 1} j_\nu \, \sqrt{-g} \, dr d\theta d\phi ,
\label{eq:lnu}
\end{equation}
\noindent where $r_{\rm g}$ converts code units to physical (CGS) units. This yields frequency-dependent luminosities $L_\nu^{\rm syn}$, $L_\nu^{\rm IC}$, and $L_\nu^{\rm tot} = L_\nu^{\rm syn} + L_\nu^{\rm IC}$.

Integration over specific energy bands provides band-limited luminosities. For the X-ray band (0.5--10 keV, corresponding to $\nu \approx 1.2 \times 10^{17}$--$2.4 \times 10^{18}$ Hz), we compute:
\begin{equation}
L_{\rm X,disk} = \int_{\nu_{\rm min}}^{\nu_{\rm max}} L_\nu^{\rm tot} \, d\nu.
\label{eq:lx_rad}
\end{equation}

\subsubsection{Angular Momentum and Velocity Structure}

The specific angular momentum is:
\begin{equation}
\lambda = -\frac{u_\phi}{u_t},
\label{eq:lambda}
\end{equation}
\noindent where $u_\phi$ and $u_t$ are components of the four-velocity. In Kerr spacetime, the Keplerian value is:
\begin{equation}
\lambda_K(r) = \frac{r^{1/2}(r^2 - 2ar^{1/2} + a^2)}{r^{3/2} - 2r^{1/2} + a}.
\label{eq:lambda_k}
\end{equation}
\noindent The ratio $\Omega/\Omega_K = \lambda/\lambda_K$ measures the degree of sub-Keplerian motion, where $\Omega = u^\phi/u^t$ is the angular velocity.

The radial velocity in the fluid frame is:
\begin{equation}
v_r = \frac{u^r}{u^t},
\label{eq:vr}
\end{equation}
\noindent which characterizes the inflow speed. The Lorentz factor characterizing outflow velocity is:
\begin{equation}
\Gamma = \alpha u^t = \frac{u^t}{\sqrt{-g^{tt}}},
\label{eq:lorentz}
\end{equation}
\noindent where $\alpha$ is the lapse function. Maximum Lorentz factor $\Gamma_{\rm max}$ in the jet region quantifies terminal jet velocities.

\section{Accretion Disk/Jet Characteristics and Implication to Observations}\label{sec:results}

\subsection{State Classification from Magnetic Flux Evolution}

The three primary simulations analyzed here are 
P2B100 (MAD state), PLB100 (INT state), and P1B100 
(SANE state), all initialized with initial 
plasma-$\beta = 100$ but with distinct magnetic field 
geometries (P2, PL, and P1, respectively) that drive 
their evolution toward qualitatively different accretion 
regimes (see \citet{raha2026} for full details of 
the parameter survey and selection criteria).

\begin{figure}[H]
\centering
\includegraphics[width=0.49\textwidth]{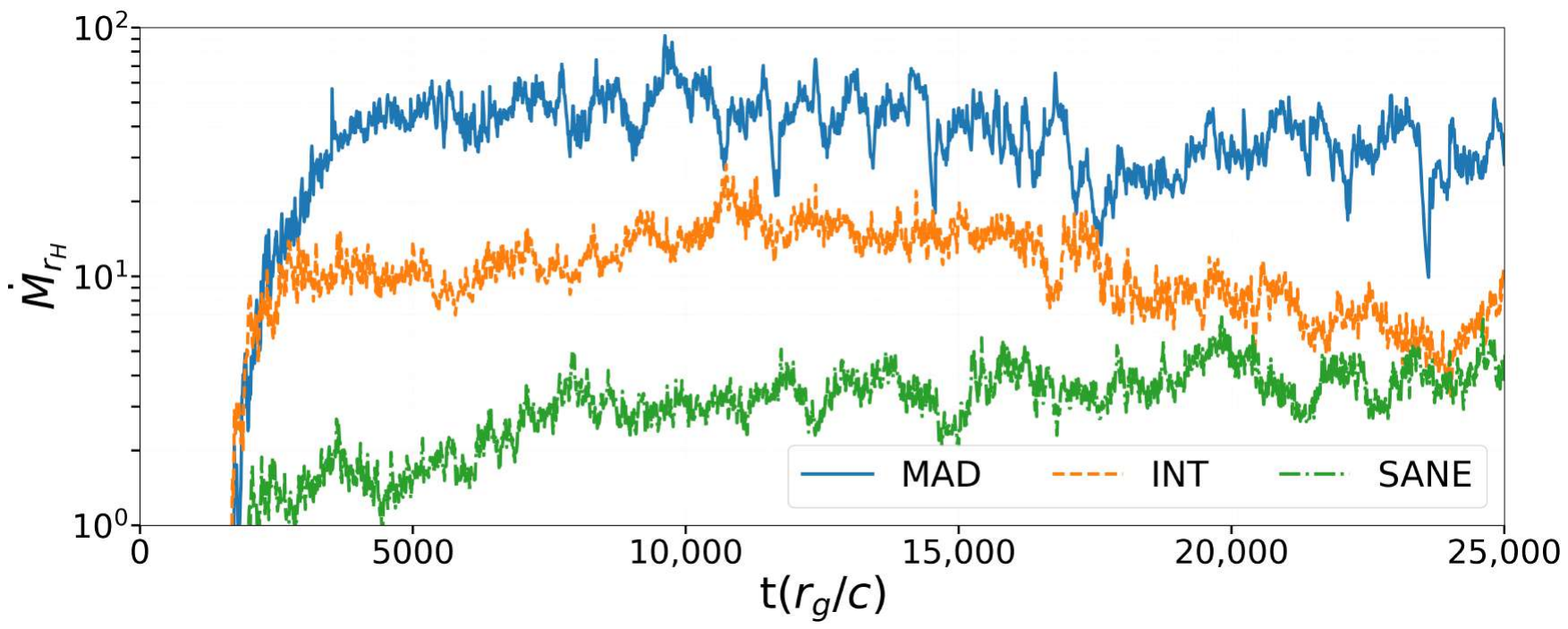}
\includegraphics[width=0.49\textwidth]{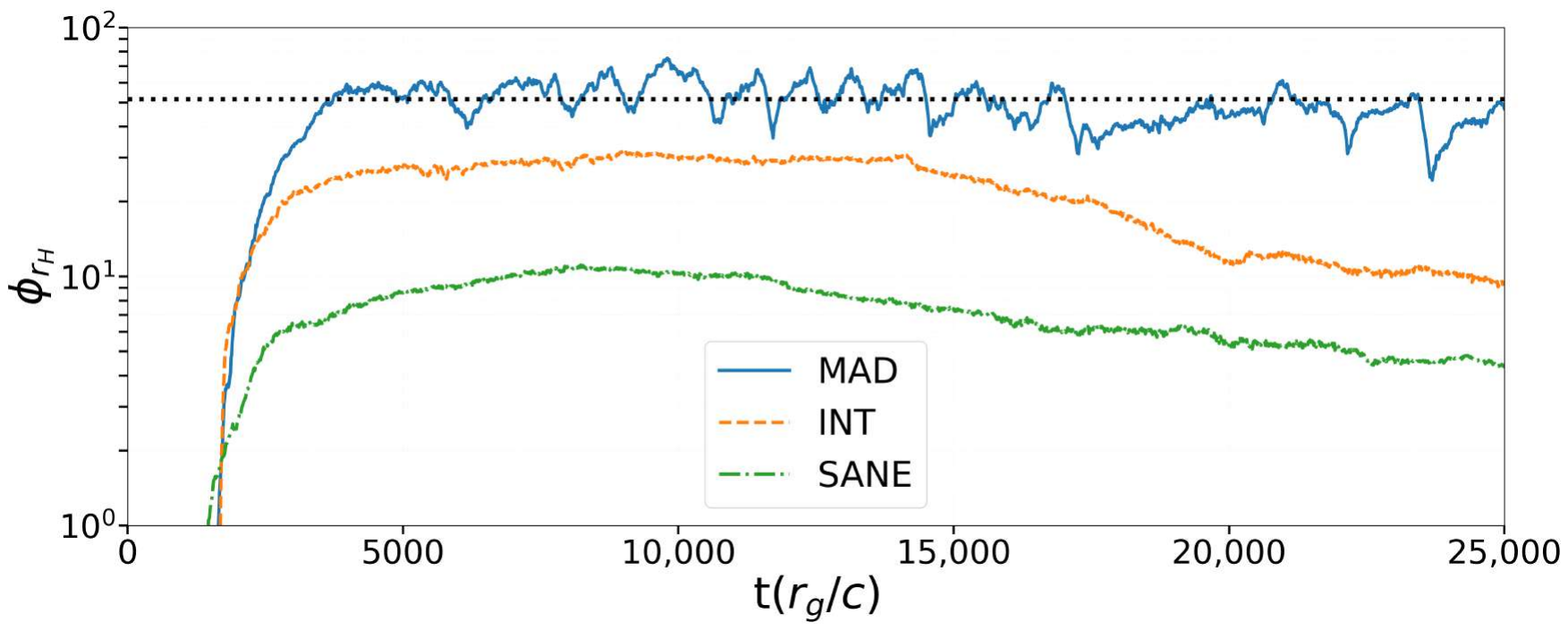}
\includegraphics[width=0.49\textwidth]{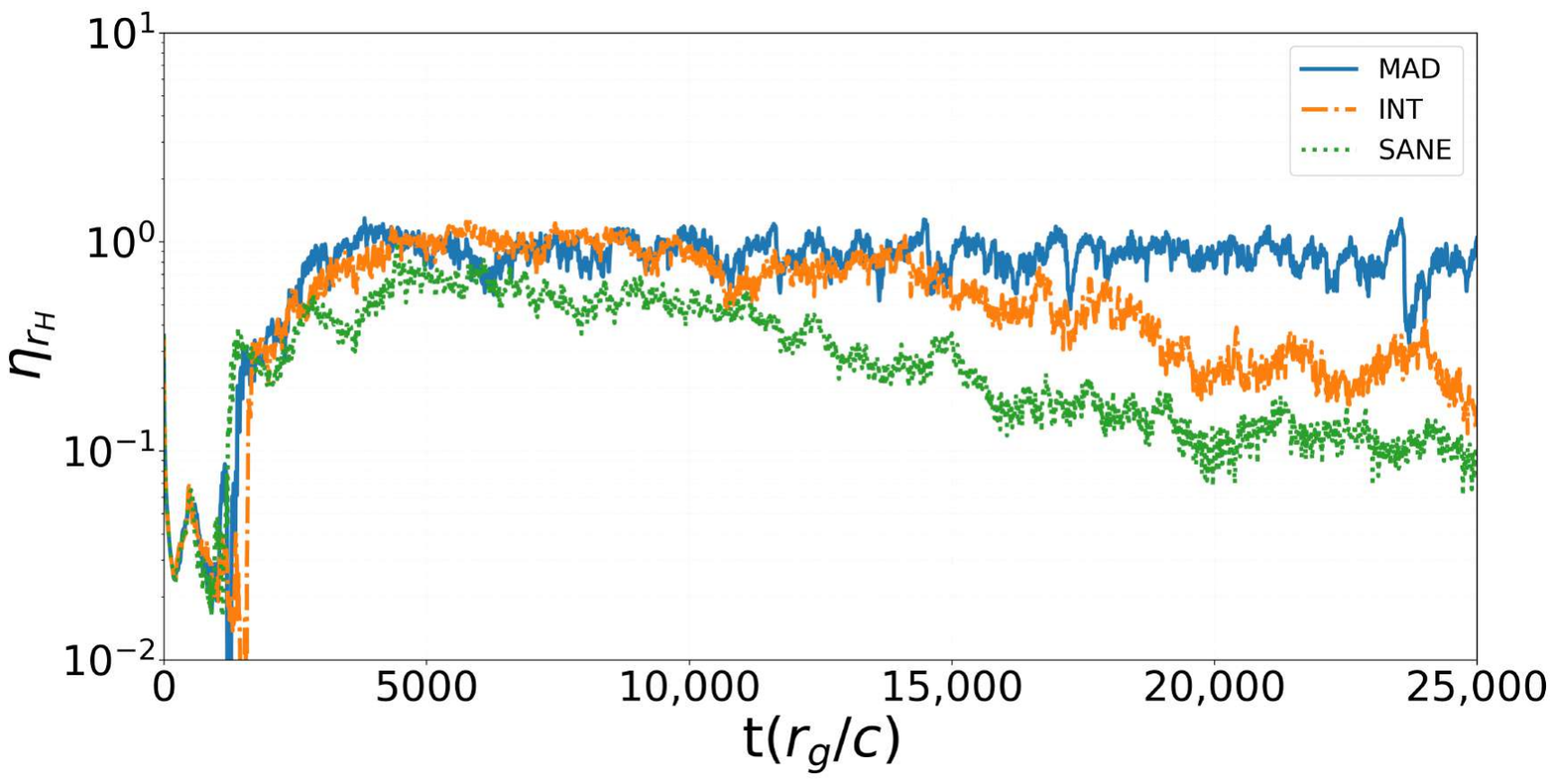}
\caption{Temporal evolution of key diagnostics for 3D simulations. \textbf{Top Left:} Mass accretion rate (in code units)  showing different variability patterns. \textbf{Top Right:} Normalized magnetic flux $\phi_{\rm BH}$ with MAD saturation threshold (dotted line at $\phi \approx 50$). \textbf{Bottom:} Jet efficiency $\eta$. MAD (P2B100, blue) exhibits quasi-periodic flux eruptions with $\langle\phi_{\rm BH}\rangle \approx 50$ and $\eta_{\rm jet} \sim 1.0\text{--}1.5$. INT (PLB100, orange) shows moderate flux $\langle\phi_{\rm BH}\rangle \sim 20\text{--}40$ with intermediate variability and $\eta_{\rm jet} \sim 0.3\text{--}0.5$. SANE (P1B100, green) maintains low flux $\langle\phi_{\rm BH}\rangle < 20$ with steady accretion and $\eta_{\rm jet} \sim 0.1\text{--}0.2$.}
\label{fig:temporal_evolution}
\end{figure}

Figure~\ref{fig:temporal_evolution} shows the temporal evolution of mass accretion rate, normalized magnetic flux $\phi_{\rm BH}$, and jet efficiency for our three primary simulations. These diagnostics reveal three distinct accretion states determined by magnetic flux accumulation.

The \textbf{MAD configuration}  reaches saturation $\phi_{\rm BH} \gtrsim 50$, triggering quasi-periodic eruptions on timescales $\Delta t \sim 1000\text{--}2000 \, r_g/c$. High variability ($\sigma_{\phi}/\mu_{\phi} \sim 0.15$ and $\sigma_{\dot{M}}/\mu_{\dot{M}} \sim 0.24$ where $\sigma_{\phi}$ and $\sigma_{\dot{M}}$ are the standard deviation of the $\phi$ and $\dot{M}$ time series, respectively, and $\mu_{\phi}$ and $\mu_{\dot{M}}$ are the mean of the same) and efficiency ($\eta_{\rm jet} > 1$) indicate black hole spin extraction via the Blandford--Znajek mechanism.

The \textbf{INT configuration}  occupies a transitional regime with $20 \lesssim \phi_{\rm BH} \lesssim 40$, exhibiting moderate variability ($\sigma_{\phi}/\mu_{\phi} \sim 0.08$ and $\sigma_{\dot{M}}/\mu_{\dot{M}} \sim 0.22$) and efficiency $\eta_{\rm jet} \sim 0.3\text{--}0.5$.

The \textbf{SANE configuration}  maintains $\phi_{\rm BH} < 20$ with minimal variability ($\sigma_{\phi}/\mu_{\phi} \sim 0.08$ and $\sigma_{\dot{M}}/\mu_{\dot{M}} \sim 0.18$) and lower efficiency $\eta_{\rm jet} \sim 0.1\text{--}0.2$, characteristic of MRI-driven turbulent transport.

\subsection{Flow Structure and Magnetization Topology}

\begin{figure}[H]
\centering
\includegraphics[width=0.49\textwidth]{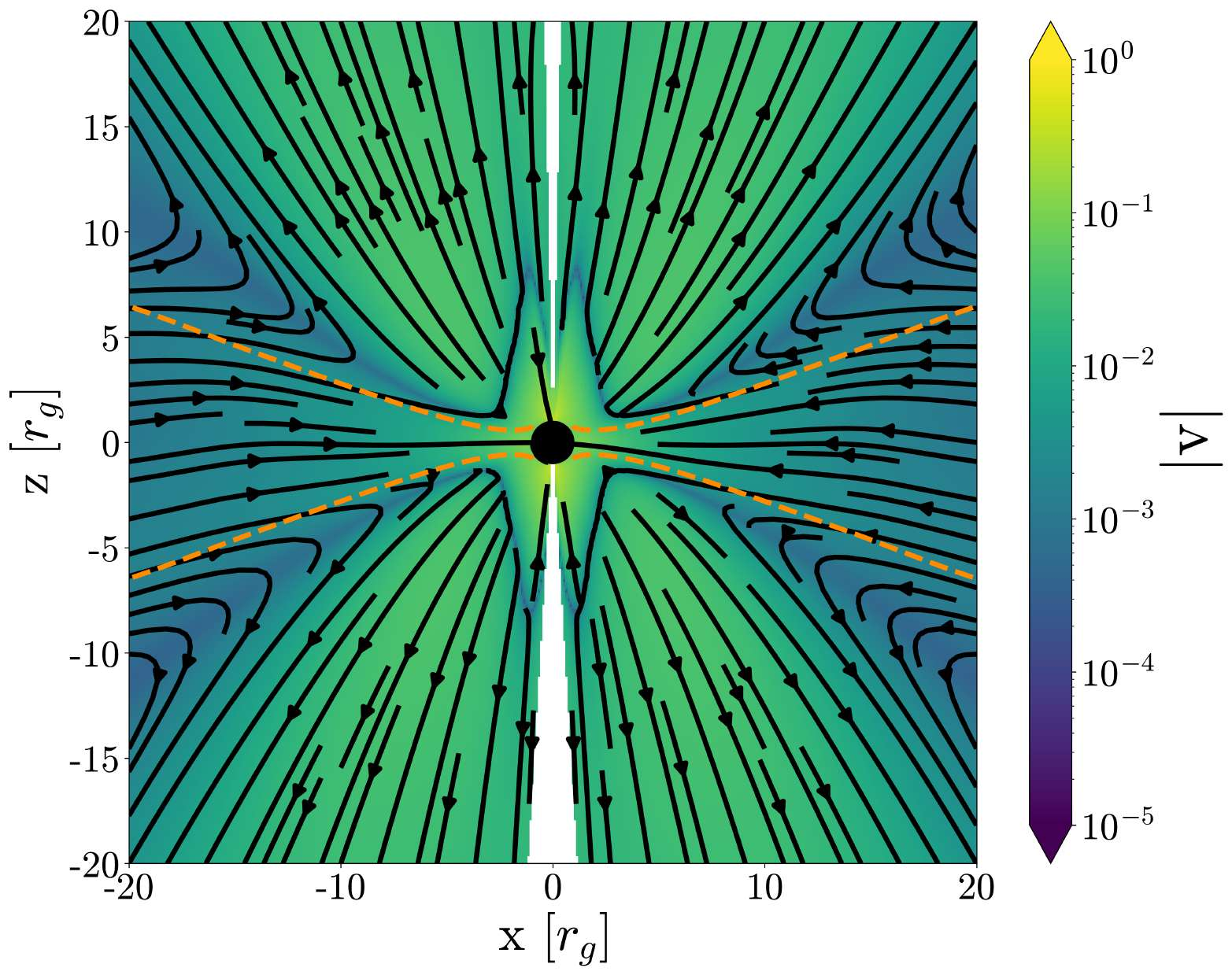}
\includegraphics[width=0.49\textwidth]{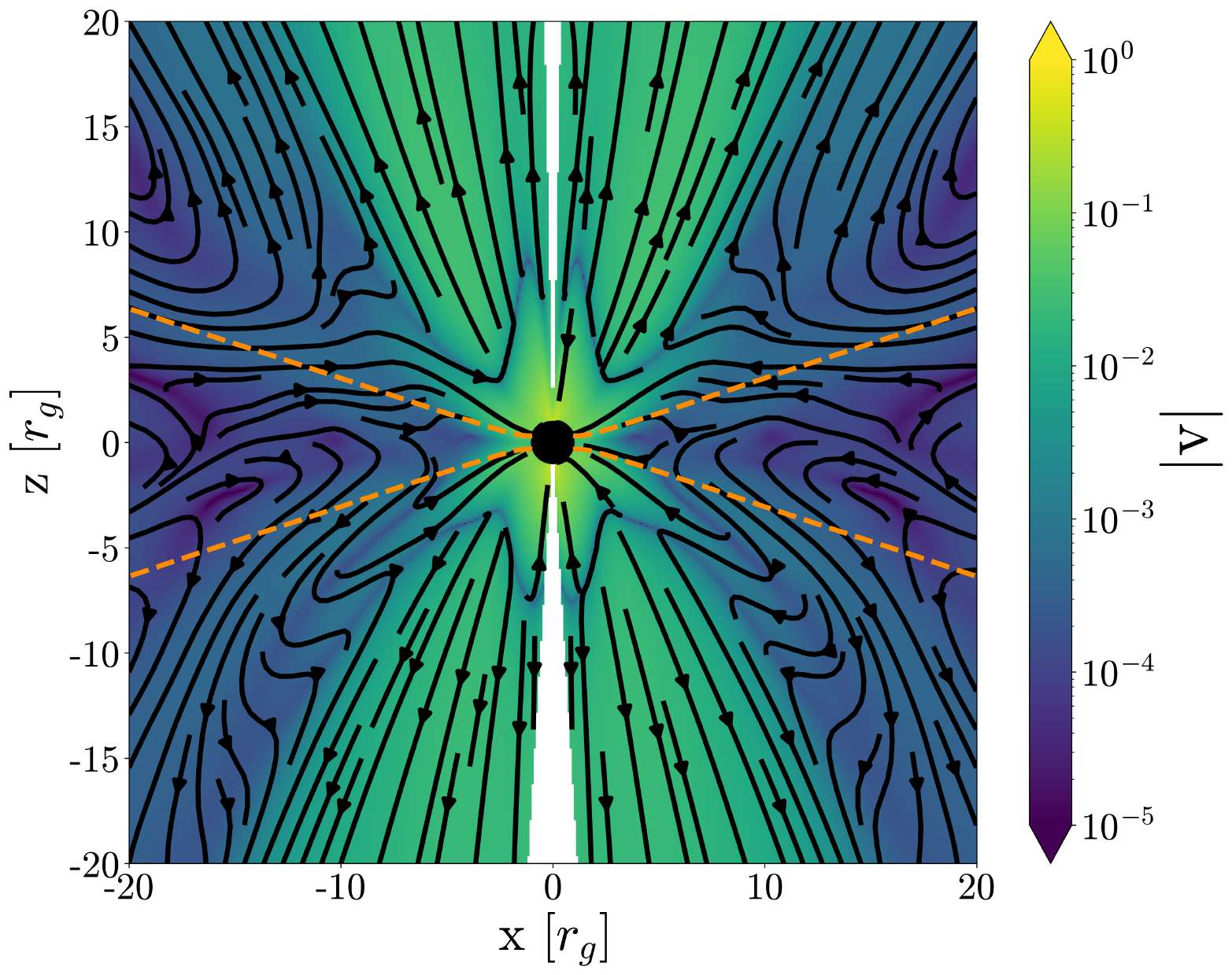}
\includegraphics[width=0.49\textwidth]{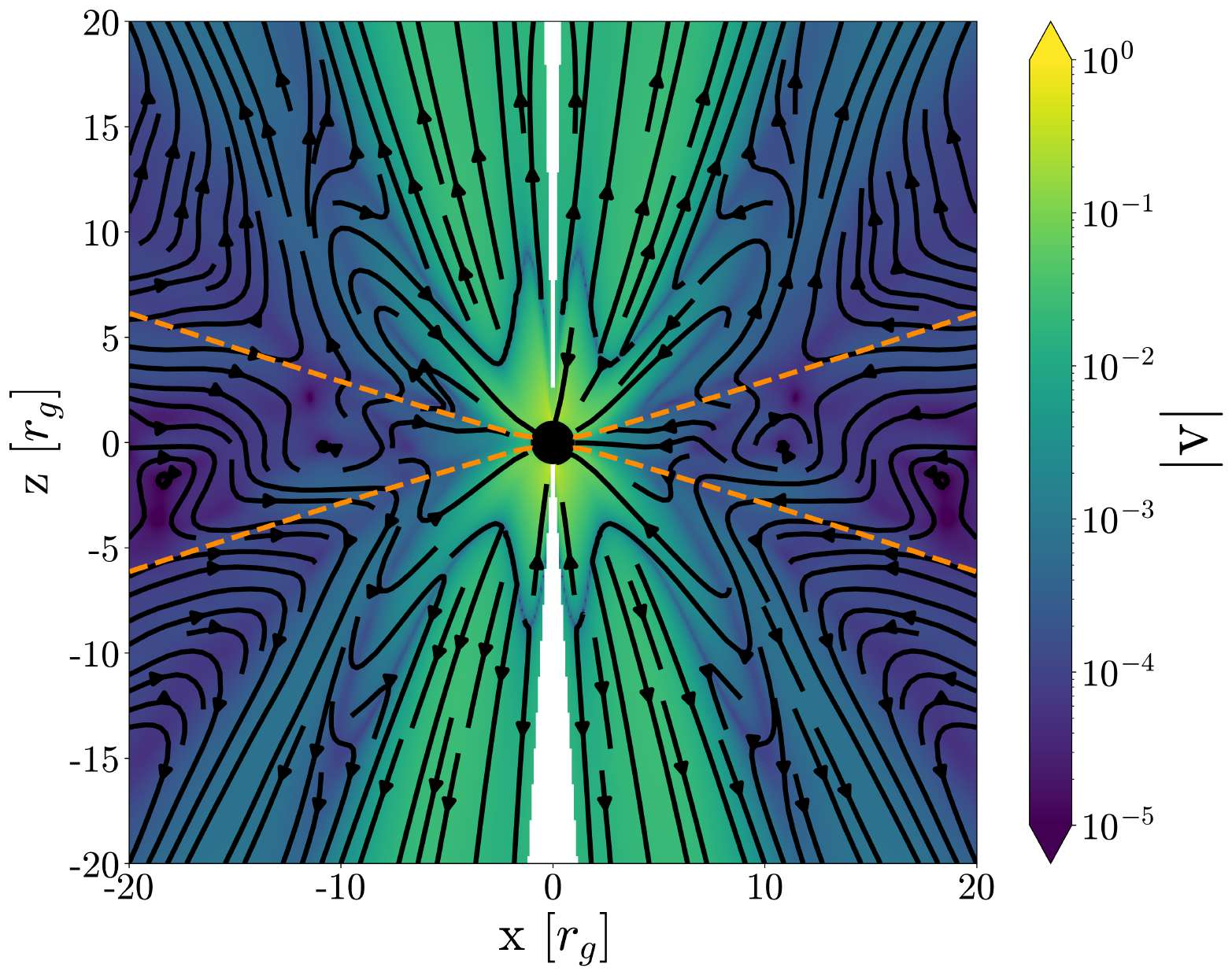}
\caption{Time-averaged velocity structure with colors showing velocity magnitude. \textbf{Top Left:} MAD exhibits highly collimated jets (opening angle$\sim 60^\circ$) with terminal velocities $v_{\rm jet} \sim 0.95c$ ($\Gamma_{\rm max} \sim 3.2$). \textbf{Top Right:} INT shows partially collimated outflows (opening angle $\sim 50^\circ$) with $v_{\rm jet} \sim 0.5c$ ($\Gamma_{\rm max} \sim 2.7$). \textbf{Bottom:} SANE displays turbulent, weakly collimated outflows (opening angle$\sim 40^\circ$) with $v_{\rm jet} \sim 0.1c$ ($\Gamma_{\rm max} \sim 2.6$). Orange dashed lines mark disk boundaries.}
\label{fig:velocity_structure}
\end{figure}

\begin{figure}[H]
\centering
\includegraphics[width=0.49\textwidth]{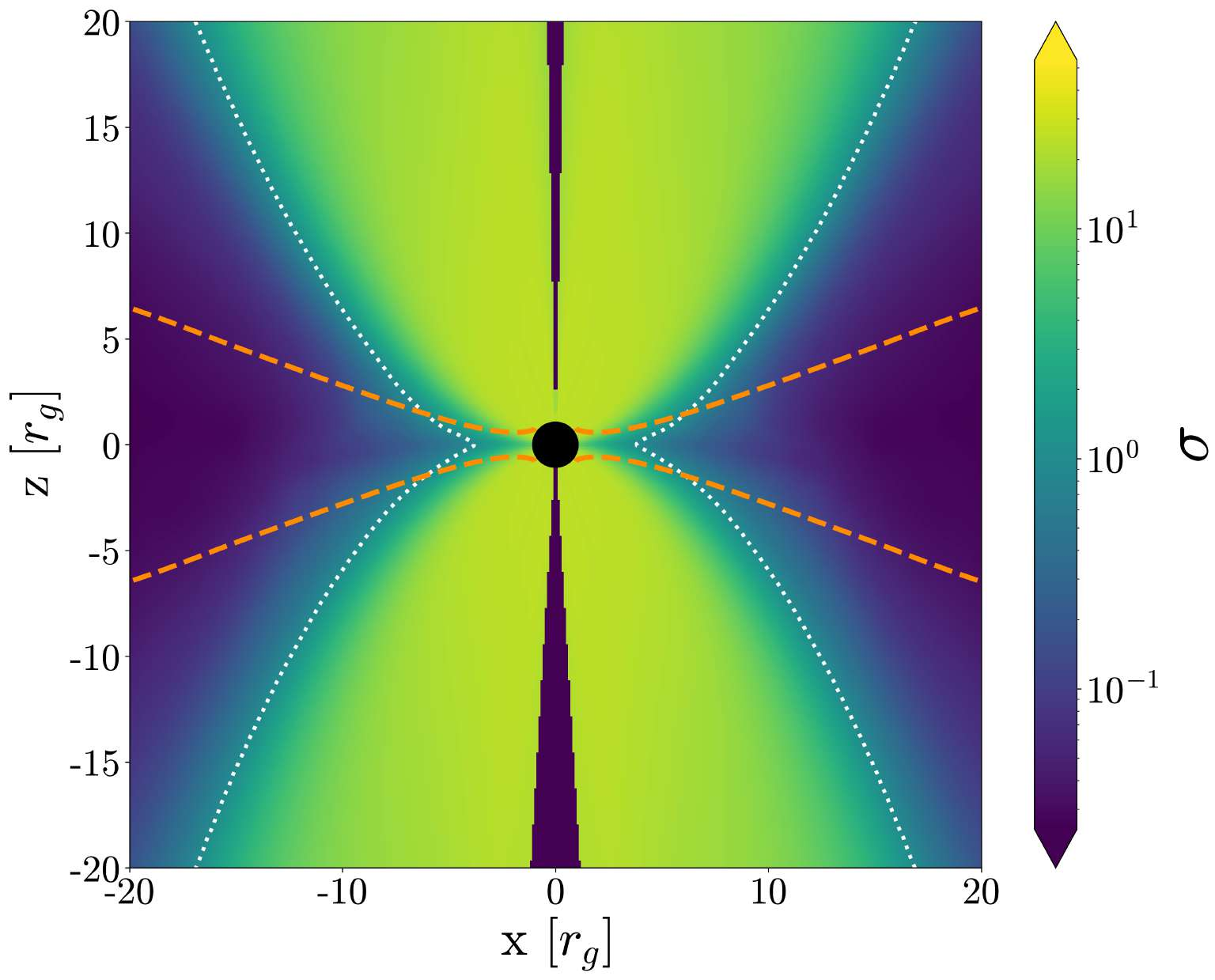}
\includegraphics[width=0.49\textwidth]{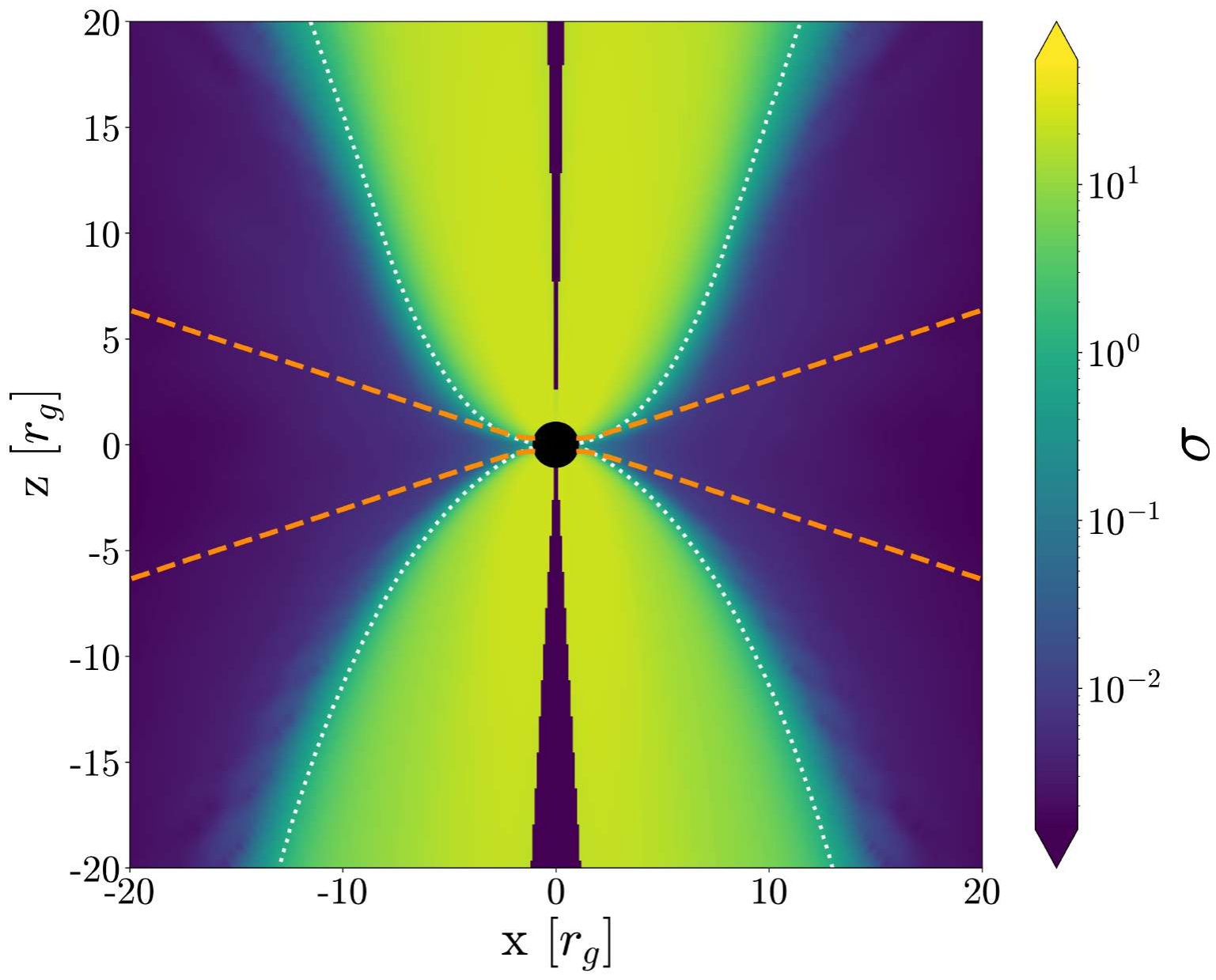}
\includegraphics[width=0.49\textwidth]{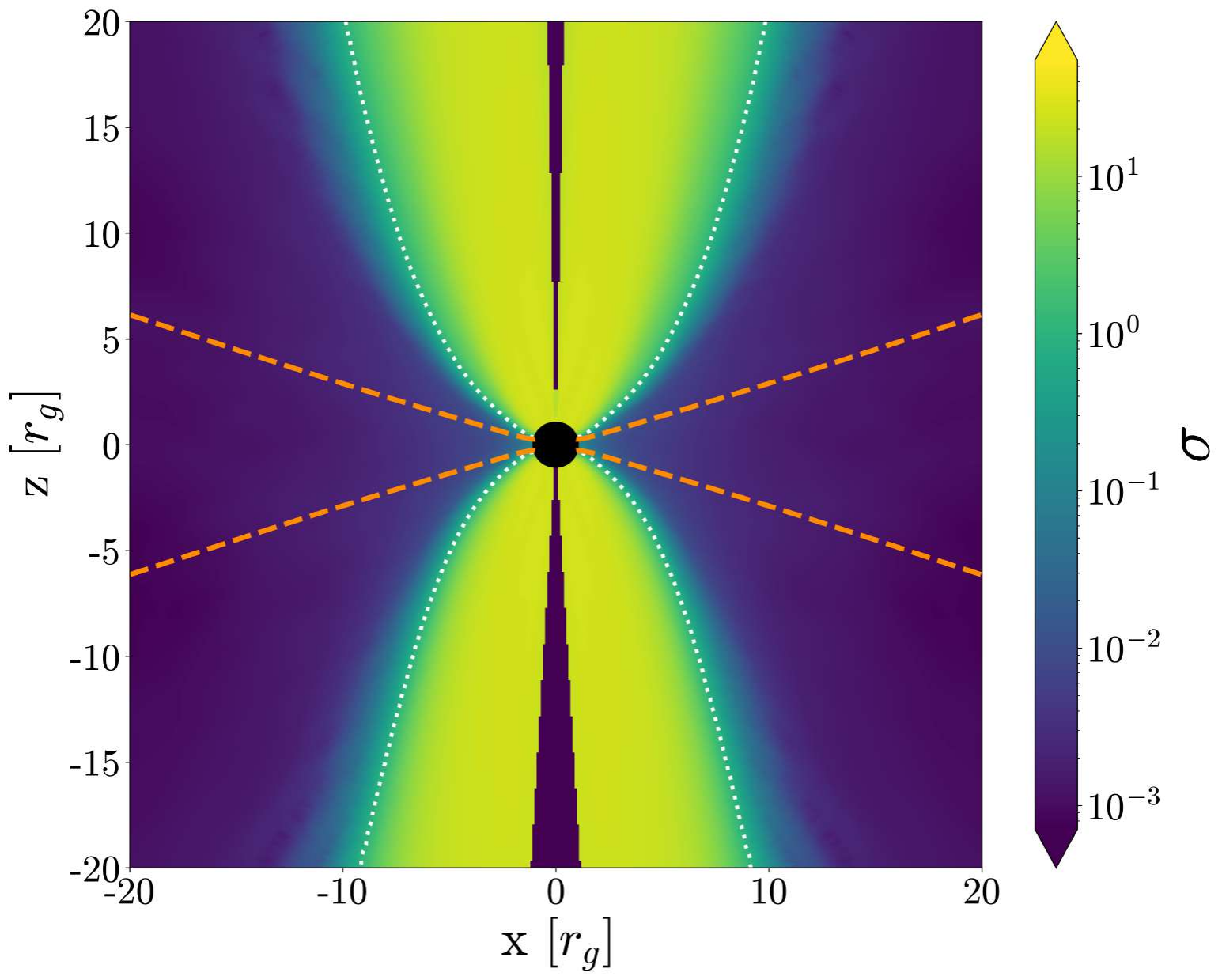}

\caption{Time-averaged
 magnetization $\sigma_m = b^2/\rho$ with the colors showing magnetization values. \textbf{Top Left:} MAD shows strong magnetic domination ($\sigma_m \sim 5\text{--}100$) in wide polar jets. \textbf{Top Right:} INT exhibits moderate magnetization ($\sigma_m \sim 0.5$) with narrower outflow regions. \textbf{Bottom:} SANE maintains weak magnetization ($\sigma_m \sim 0.1$) throughout. White dotted contours mark $\sigma_m = 1$ (jet-disk boundary); orange dashed lines show disk scale height.}
\label{fig:magnetization_structure}
\end{figure}

Figures~\ref{fig:velocity_structure} and \ref{fig:magnetization_structure} display the time-averaged velocity fields and magnetization parameter $\sigma_m = b^2/\rho$ in the meridional plane, revealing distinct flow topologies.

The magnetization structure naturally partitions each flow into three regions: magnetically dominated jets ($\sigma_m > 1$), gas-pressure dominated disks ($\sigma_m < 1$ within $h/r$), and intermediate winds ($\sigma_m < 1$ outside $h/r$). MAD maintains $\langle\sigma_m\rangle_\rho \sim 5\text{--}6$ at the horizon with extensive jet regions; INT achieves $\langle\sigma_m\rangle_\rho \sim 0.5$ with mixed topology; SANE shows $\langle\sigma_m\rangle_\rho \sim 0.1$ with minimal jet formation. See Figure~\ref{fig:magnetization_structure} for clarity.

\subsection{Energy Extraction: Jets Versus Disk/Wind}

\begin{figure}[H]
\centering
\includegraphics[width=0.49\textwidth]{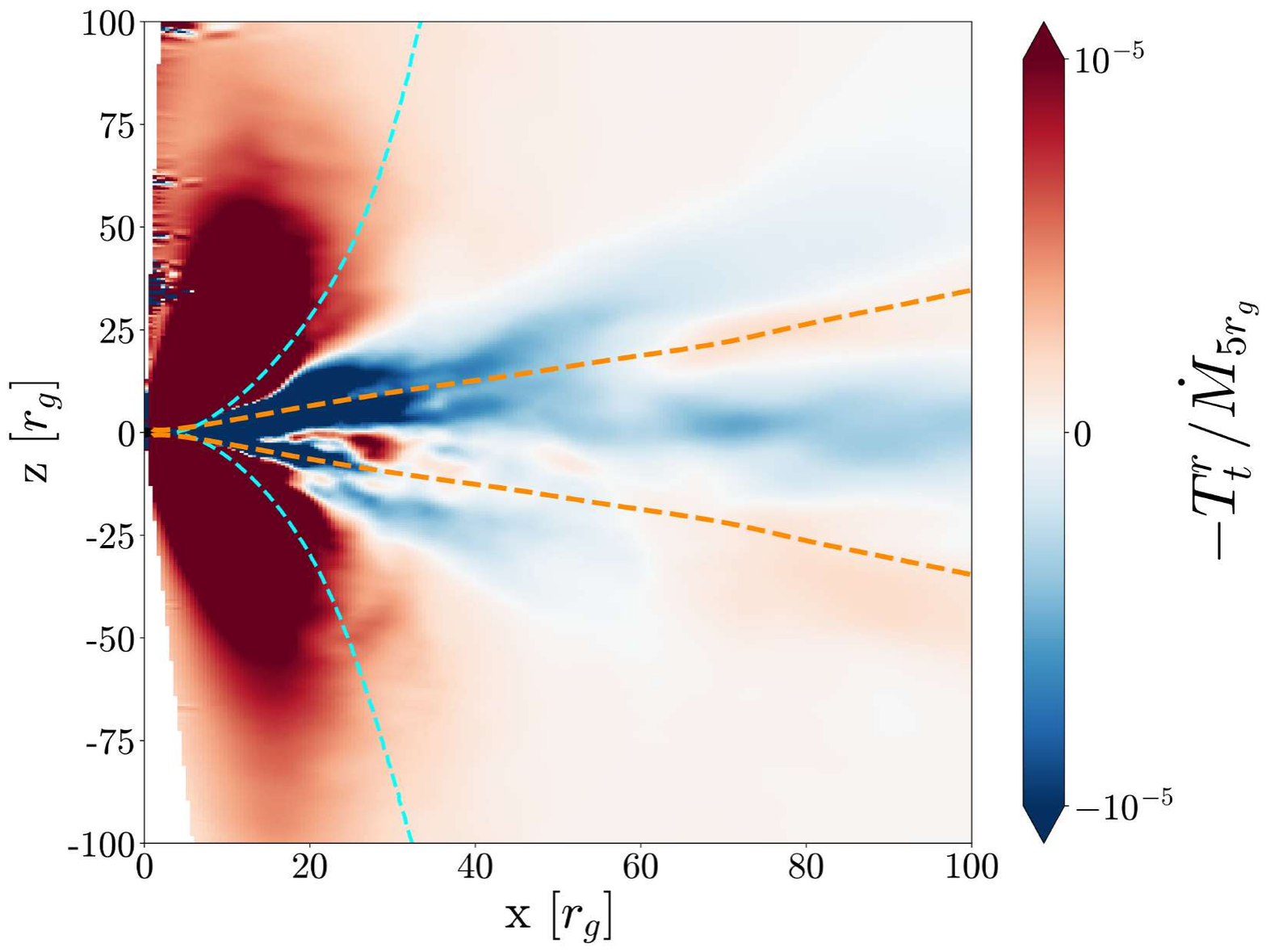}
\includegraphics[width=0.49\textwidth]{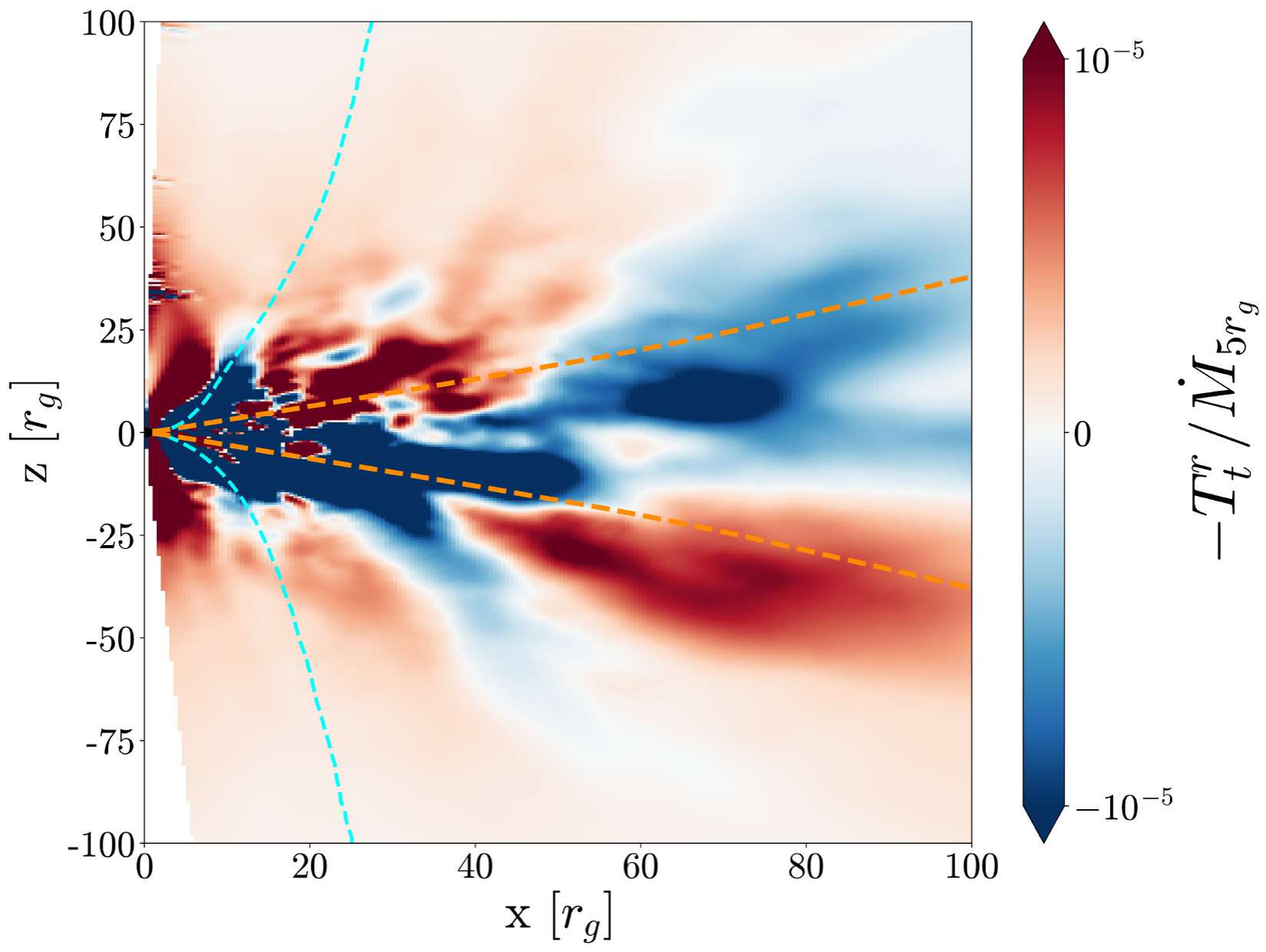}
\includegraphics[width=0.49\textwidth]{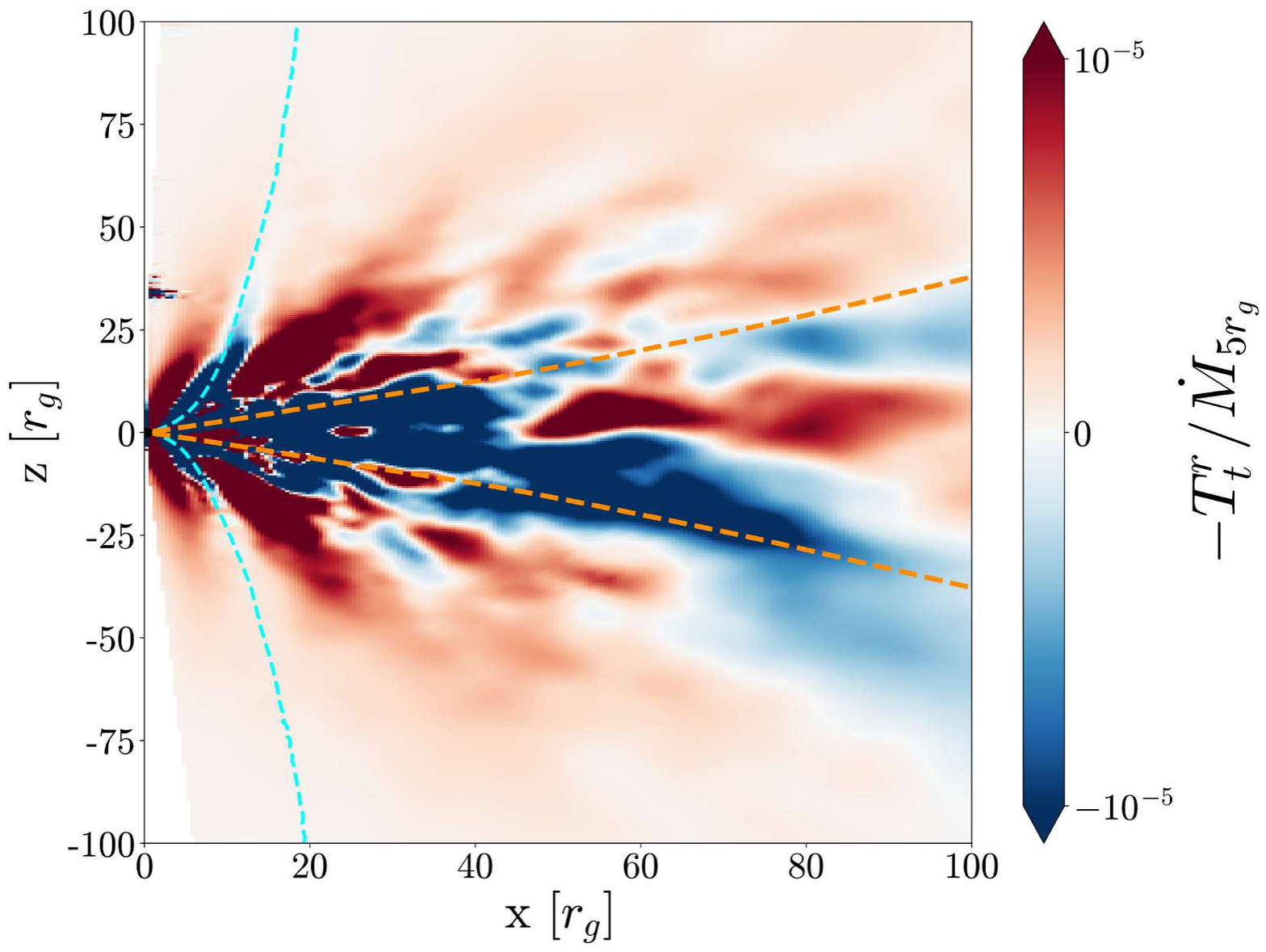}
\caption{Normalized
 radial energy flux $-T^r_{\phantom{r}t}/(\dot{M}_{5r_g} c^2)$. Positive values (red) indicate outward energy transport; negative values (blue) show inward advection. The orange dashed lines represent the boundary of the disk region and the cyan dashed lines are the boundaries of the jet region. \textbf{Top Left:} MAD concentrates energy extraction in polar jets through electromagnetic Poynting flux. \textbf{Top Right:} INT shows mixed jet and wind contributions. \textbf{Bottom:} SANE is dominated by thermally driven disk/wind energy transport.}
\label{fig:stress_energy_flux}
\end{figure}

Figure~\ref{fig:stress_energy_flux} displays the spatial distribution of energy transport. It reveals fundamental differences in energy extraction mechanisms across magnetic states. In the MAD configuration (Figure~\ref{fig:stress_energy_flux}, top left), outward energy flux (red regions) is strongly concentrated within polar funnels bounded by the cyan dashed jet boundaries, indicating highly collimated electromagnetic Poynting flux extraction. The disk region (within orange dashed boundaries) shows predominantly inward energy advection (blue), with minimal outward transport. In contrast, the SANE configuration (Figure~\ref{fig:stress_energy_flux}, bottom) displays broadly distributed outward energy transport extending well beyond the disk scale height, with significant flux in both equatorial and polar regions, indicating gas pressure driven winds dominating over organized jets.  The INT state occupies an intermediate regime, with Figure~\ref{fig:stress_energy_flux} (top right) showing partially collimated outflows alongside substantial disk/wind transport. MAD extracts energy electromagnetically through organized jets, SANE dissipates energy through disk turbulence and winds, while INT maintains balanced electromagnetic and gas pressure channels.

\subsection{Spectral Energy Distributions}
\label{sec:sed_results}

\begin{figure}[H]
\includegraphics[width=0.73\textwidth]{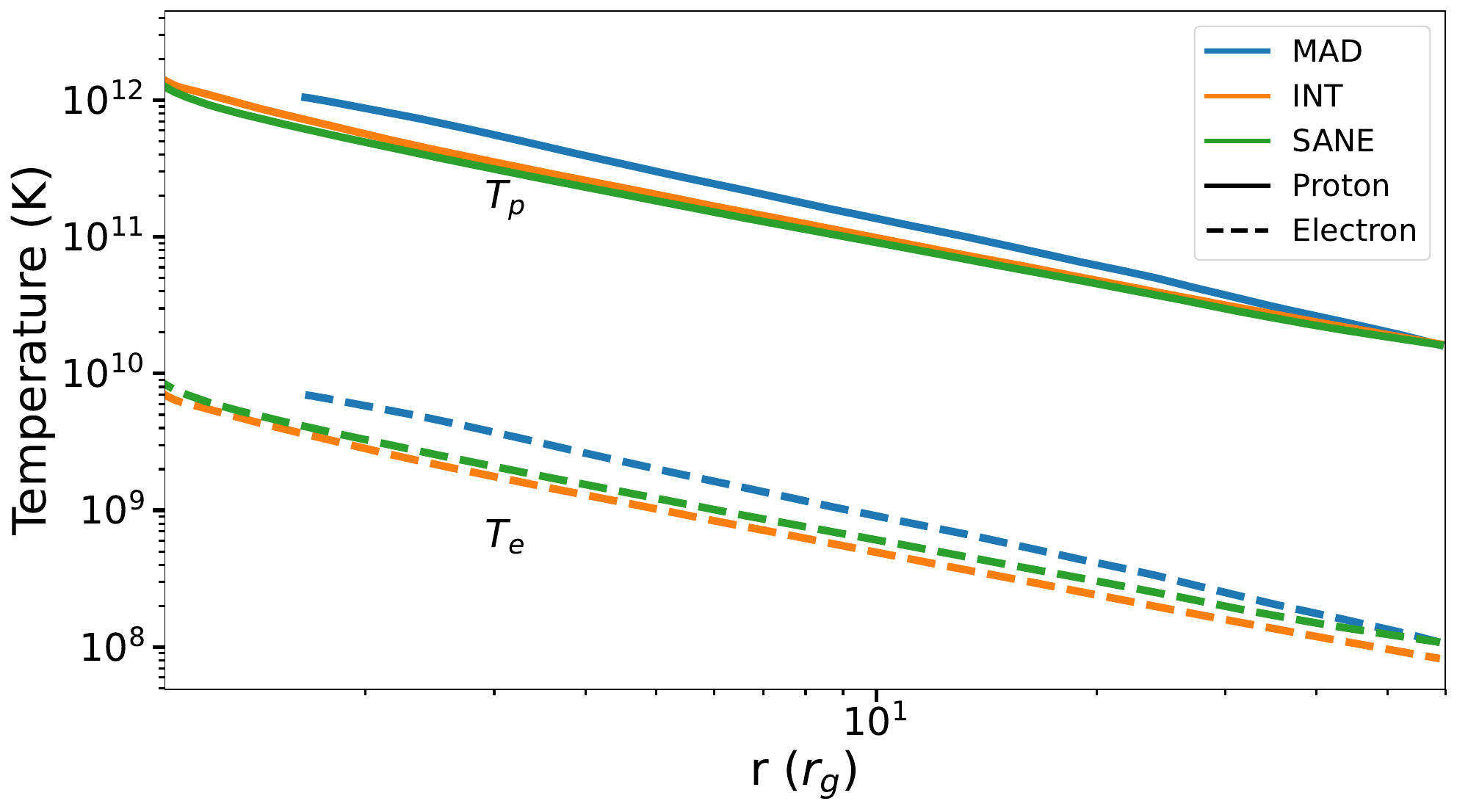}
\caption{Density-weighted radial profiles of proton temperature $T_p$ (solid
lines) and electron temperature $T_e$ (dashed lines) for MAD (blue), INT
(orange), and SANE (green) states, scaled to GRS~1915+105 ($M = 14\,M_\odot$,
$\dot{m} = 0.01$). The state-dependent proton-to-electron temperature ratios
$R = T_p/T_e$ of 150 (MAD), 200 (INT), and 150 (SANE) are
chosen such that the electron temperature profiles closely match across all
three states in the inner disk region (and $T_e$ is the same at the inner edge). This ensures that spectral differences
between states arise from their distinct magnetic field structures and dynamics
rather than from differences in electron heating. Note that for MAD the
disk region starts from $r\sim 1.65 r_g$ as shown in
Figure~\ref{fig:magnetization_structure}, as the inner region is highly
magnetized. Hence, the temperatures are shown from that radius.}
\label{fig:temp_profile}
\end{figure}

Before presenting the spectra, we justify the choice of $R$ values.
Figure~\ref{fig:temp_profile} shows the density-weighted, radially resolved
proton and electron temperature profiles for all three magnetic states, scaled
to GRS~1915+105 parameters ($M = 14\,M_\odot$, $\dot{m} = 0.01$). The
state-dependent $R$ values are chosen such that the resulting electron
temperature profiles are closely matched across MAD, INT, and SANE
configurations throughout the disk region ($r \lesssim 60\,r_g$). Any
differences in the emergent spectra and luminosities arise purely from the
underlying dynamics and magnetic field structure---specifically the
magnetization topology, plasma-$\beta$, and optical depth---rather than from
differences in electron heating. The proton temperatures, by contrast, differ
between states, reflecting the distinct magnetic energy dissipation rates in
each configuration. This approach allows us to isolate the role of magnetic
flux geometry in shaping the radiative output, which is the central objective
of this study.

\begin{figure}[H]
\includegraphics[width=0.74\textwidth]{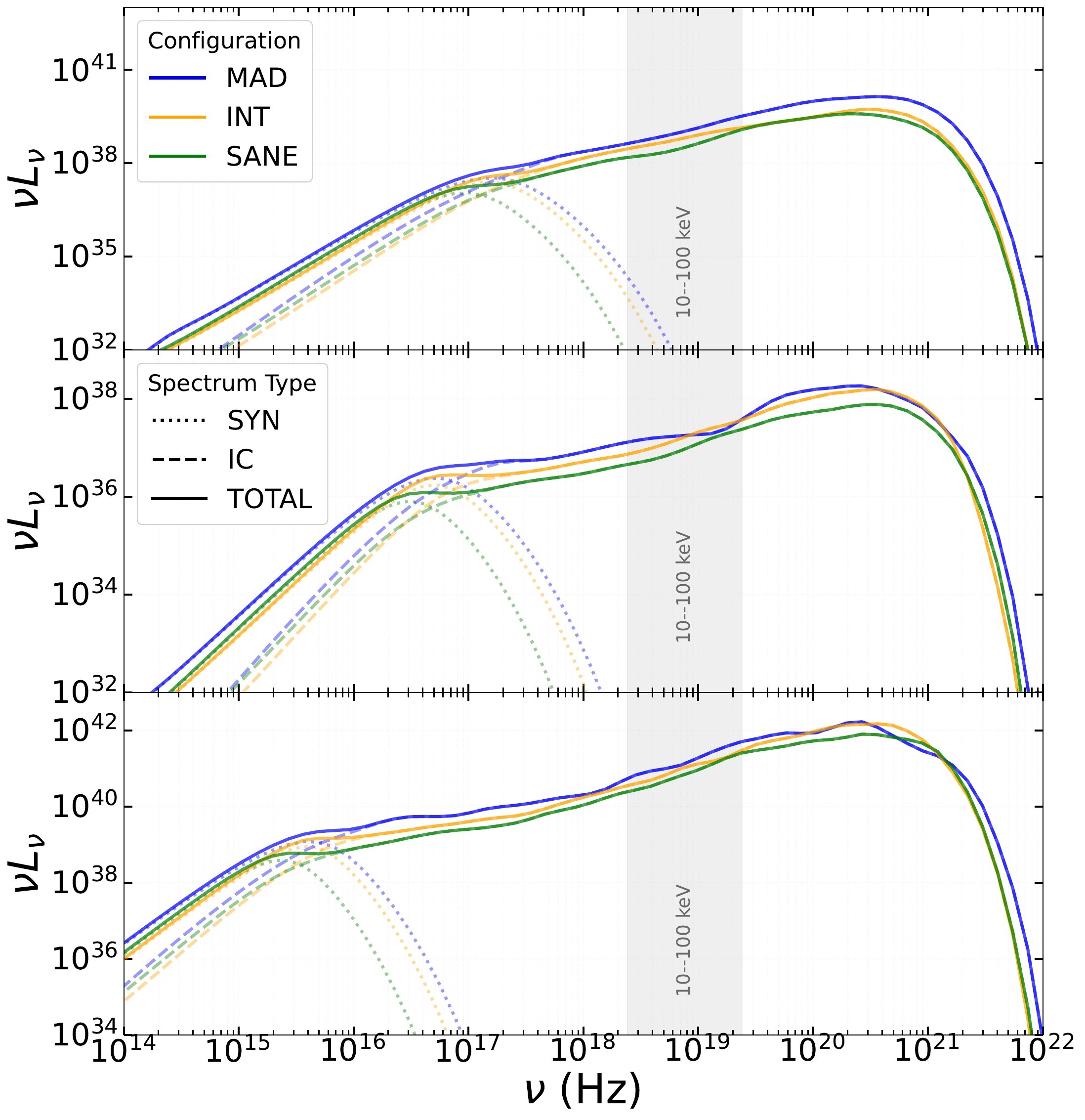}
\caption{Broadband
 spectral energy distributions for three black hole systems
spanning $10$--$10^4 \, M_\odot$. \textbf{Top:} GRS~1915+105 ($M = 14 \, M_\odot$, $\dot{m} = 0.01$). \textbf{Middle:} Cyg~X-1 ($M = 20 \, M_\odot$, $\dot{m} = 0.002$). \textbf{Bottom:} HLX-1 ($M = 2 \times 10^4 \, M_\odot$, $\dot{m} = 0.003$). Solid lines show total emission (synchrotron + inverse-Compton), dotted lines show the synchrotron component, and dashed lines show
the inverse-Compton component. Colors indicate magnetic state: MAD (blue), INT (orange), SANE (green). All spectra exhibit inverse-Compton peaks at $\nu \sim 2 \times 10^{20}$--$4 \times 10^{20}$~Hz; the luminosity hierarchy
among magnetic states depends on source mass and accretion rate. The shaded grey region in each panel marks the hard X-ray band (10--100~keV, $\nu \approx 2.4 \times 10^{18}$--$2.4 \times 10^{19}$~Hz) discussed in
Section~\ref{sec:lum_hierarchy}.}
\label{fig:spectra_combined}
\end{figure}

Figure~\ref{fig:spectra_combined} displays broadband spectral energy
distributions calculated from time-averaged simulation data for the three
sources, decomposed into synchrotron (SYN, dotted) and inverse-Compton (IC,
dashed) components, and total (solid). These provide the radiative signatures
corresponding to the kinematic energy fluxes quantified in
Section~\ref{sec:diagnostics}.

All magnetic states exhibit inverse-Compton-dominated spectra with a
characteristic double-peaked structure. Synchrotron emission extends from
radio through optical and ultraviolet frequencies, peaking at $\nu \sim
10^{15}$--$10^{17}$~Hz, while inverse-Compton dominates at higher energies
with peaks at $\nu \sim 2 \times 10^{20}$--$4 \times 10^{20}$~Hz
corresponding to photon energies $\sim 0.8$--$1.5$~MeV.

\vspace{-9pt}

The X-ray band (0.1--100 keV, $\nu \approx 1 \times 10^{17}$--$3 \times
10^{19}$ Hz) spans from the low-energy synchrotron tail through the rising
inverse-Compton component. In this band, emission transitions from
synchrotron-dominated at soft energies to inverse-Compton dominated at hard energies,
with the crossover occurring at $\sim $10--$30$ keV depending on magnetic
state and source parameters. The peak electron temperature in the hot
inner disk, which dominates the inverse-Compton emissivity, reaches
$\Theta_{e,\mathrm{peak}} \sim 1.5$ ($T_e \sim 9 \times 10^{9}$~K), placing
the inverse-Comptonization spectral cutoff at $\sim 2 k_B T_{e,\mathrm{peak}} \approx
1.5$~MeV --- consistent with the inverse-Compton peaks at $\nu \sim 2 \times
10^{20}$--$4 \times 10^{20}$~Hz in Figure~\ref{fig:spectra_combined}. It
follows the same $2k_BT_e$ scaling demonstrated by \citet{Takahara1981}
through Monte Carlo modelling of unsaturated inverse-Comptonization in hot plasma.

\subsection{Luminosity Hierarchy, Scaling and Implications to Black Hole X-Ray Binaries}
\label{sec:lum_hierarchy}

Figure~\ref{fig:spectra_combined} displays the broadband SEDs for all three
sources and magnetic states. The hard X-ray band (10--100~keV, corresponding
to $\nu \approx 2.4 \times 10^{18}$--$2.4 \times 10^{19}$~Hz) is indicated
by the shaded grey region in each panel of Figure~\ref{fig:spectra_combined}.
In this band, the radiative output follows MAD $>$ INT $>$ SANE. For the
parameters suitable for GRS~1915+105, the overall inverse-Compton spectral
peaks are $\nu L_\nu \sim 1.4 \times 10^{40}$ (MAD), $5.3 \times
10^{39}$ (INT), and \mbox{$3.8 \times 10^{39}$ erg~s$^{-1}$} (SANE), yielding a
MAD/SANE ratio of $\sim 3.5$. Within the hard X-ray band (10--100 keV),
simulated luminosities span $\nu L_\nu \sim 4.8 \times 10^{38}$--$3.2
\times 10^{39}$ erg~s$^{-1}$ for MAD, $3.2 \times 10^{38}$--$1.4 \times
10^{39}$ erg~s$^{-1}$ for INT, and $1.6 \times 10^{38}$--$1.2 \times
10^{39}$ erg~s$^{-1}$ for SANE states. The MAD/SANE ratio is $\sim 3$ in the
hard X-ray band.

This hierarchy is driven by magnetic field strength rather than electron
temperature. As demonstrated in Figure~\ref{fig:temp_profile}, the electron
temperature profiles are closely matched across all three states by
construction through the choice of state-dependent $R$ values. The spectral
differences, therefore, arise from the distinct magnetic field structures of
each accretion state. MAD configurations sustain the strongest large-scale
magnetic fields near the horizon, producing synchrotron emission, for
GRS~1915+105 like parameters, \mbox{$\sim $3 times} more luminous than SANE
($\nu L_{\nu,\rm syn}^{\rm MAD} \sim 3.3 \times 10^{37}$ versus $\sim 1.1
\times 10^{37}$ erg~s$^{-1}$) and at higher characteristic synchrotron
frequencies. This abundance of high-energy seed photons in MAD drives a
correspondingly stronger inverse-Compton component which
exceeds SANE by a factor $\sim 3.5$. The inverse-Compton spectral peak
frequency follows the same \mbox{MAD $>$ INT $>$ SANE} ordering as the luminosity
hierarchy. For GRS~1915+105, MAD peaks at $\nu_{\rm IC}^{\rm peak} \sim
3.6 \times 10^{20}$~Hz ($\sim $1.5~MeV), INT at $\sim 2.7 \times 10^{20}$~Hz
($\sim $1.1~MeV), and SANE at $\sim 2.0 \times 10^{20}$~Hz ($\sim
$0.8~MeV). This ordering reflects the magnetic field hierarchy: MAD's
stronger large-scale magnetic fields produce higher characteristic synchrotron
frequencies, supplying more energetic seed photons for inverse-Compton upscattering
and thereby shifting the inverse-Compton spectral peak to higher frequencies.
The weaker seed photon field in SANE consequently results in both a lower
inverse-Compton peak frequency and a substantially lower overall
inverse-Compton power.

Scaling to Cyg~X-1 and HLX-1 shifts absolute luminosities according to $L
\propto \dot{m} M$. The set of parameters corresponding to Cyg~X-1 exhibits
luminosities in the hard X-ray band (10--100 keV): $\nu L_\nu \sim 1.4
\times 10^{37}$--$3.8 \times 10^{37}$ erg~s$^{-1}$ (MAD), $8.1 \times
10^{36}$--$3.6 \times 10^{37}$ erg~s$^{-1}$ (INT), and $4.9 \times
10^{36}$--$2.4 \times 10^{37}$ erg~s$^{-1}$ (SANE), appropriate for
$\dot{m} = 0.002$ of Cyg~X-1. Similarly, the parameters corresponding to
HLX-1 reach $\nu L_\nu \sim 6.8 \times 10^{40}$--$5.1 \times 10^{41}$
erg~s$^{-1}$ (MAD), $4.0 \times 10^{40}$--$3.0 \times 10^{41}$
erg~s$^{-1}$ (INT), and $2.7 \times 10^{40}$--$2.6 \times 10^{41}$
erg~s$^{-1}$ (SANE) in the hard X-ray band, owing to its higher black hole
mass. The preserved spectral characteristics across $\sim 10^3$ range in mass
suggest universal emission physics governed by dimensionless\mbox{ parameters.}

Based on the above spectral descriptions for MAD, INT and SANE, we estimate
the following powers of the sources under consideration. For GRS~1915+105
($M = 14\,M_\odot$, \mbox{$\dot{m} = 0.01$)}, which appears to be a combination of
all three states, $P_{\rm jet}=2\times10^{38}$ erg s$^{-1}$, \mbox{$7\times10^{37}$
erg s$^{-1}$}, $2\times10^{37}$ erg s$^{-1}$ and $L_{\rm
X,disk}=3.2\times10^{39}$ erg s$^{-1}$, $1.4\times10^{39}$ erg s$^{-1}$,
$1.2\times10^{39}$ erg s$^{-1}$, respectively, for MAD, INT and SANE. Scaling
to Cyg~X-1 ($M = 20\,M_\odot$, $\dot{m} = 0.002$), which is predicted to have an INT
configuration, we obtain $P_{\rm jet} \sim 8 \times 10^{36}$ erg s$^{-1}$
and $L_{\rm X, disk} \sim 3.6 \times 10^{37}$ erg s$^{-1}$ (based on
Figure~\ref{fig:spectra_combined}), matching its sustained hard state. For
HLX-1 ($M = 2 \times 10^4\,M_\odot$, $\dot{m} = 0.003$) with the MAD
configuration, we predict \mbox{$P_{\rm jet} \sim 3 \times 10^{41}$ erg s$^{-1}$}
and $L_{\rm X, disk} \sim 5.1 \times 10^{41}$ erg s$^{-1}$, explaining
its apparent high luminosity through efficient electromagnetic extraction (see
Figure~\ref{fig:spectra_combined}).

\section{Implications to GRS~1915+105 and Comparing with Other Sources}
\label{sec:grs_all}
\subsection{Identifying GRS 1915+105: Twelve Classes from Three Accretion States}
\label{sec:grs_classes}

\textls[-15]{The twelve temporal classes of GRS 1915+105 \citep{belloni2000}, with
distinctive spectral properties\mbox{ \citep{adegoke2018}}}, map onto our magnetic
states using $M = 14\,M_\odot$ and $\dot{m} = 0.01$.

\textbf{MAD classes}  ($\chi$, $\rho$, $\alpha$): These hardest classes exhibit
72--77\% power-law spectral dominance \citep{adegoke2018}, indicating
jet-dominated emission. The $\chi$ class maintains persistent radio jets with
steady hard spectra. Our simulations predict $P_{\rm jet} \sim 2 \times
10^{38}$ erg s$^{-1}$ (Table~\ref{tab:state_properties}) and $L_{\rm X,
disk, max} \sim 3.2 \times 10^{39}$ erg s$^{-1}$
(Table~\ref{tab:state_properties}, Figure~\ref{fig:spectra_combined}), with
$L_{\rm X, disk}/P_{\rm jet} \sim 16$, matching observed hard state
luminosities. The sub-Keplerian rotation ($\Omega/\Omega_K \sim 0.5$) and
high magnetization ($\sigma_m \sim 6$) at the horizon enable efficient
Blandford--Znajek extraction with $\eta_{\rm jet} > 1$.

\textbf{INT classes}  ($\nu$, $\kappa$, $\beta$, $\theta$): These transitional
classes show balanced thermal and non-thermal spectral components, often
exhibiting rapid switches between hard and soft states on timescales of
seconds to minutes. The variability arises because plasma-$\beta \sim 1$
(where magnetic and gas pressure are comparable) represents a critical
unstable point---small perturbations in either pressure can temporarily shift
local regions toward more MAD-like or SANE-like behavior, driving observed
spectral oscillations. The $\beta$ class exemplifies this instability with
near-equal diskbb and power-law contributions. Our simulations predict $P_{\rm
jet} \sim 7 \times 10^{37}$ erg s$^{-1}$ (see Table~\ref{tab:state_properties})
and $L_{\rm X, disk, max} \sim 1.4 \times 10^{39}$ erg s$^{-1}$ (see
Figure~\ref{fig:spectra_combined}, Table~\ref{tab:state_properties}),
reflecting strong X-ray dominance ($L_{\rm X, disk}/P_{\rm jet} \sim 20$)
with episodic rather than steady jet activity. The moderate magnetization
($\sigma_m \sim 0.5$) produces partially collimated jets coexisting with
substantial disk winds.

\textbf{SANE classes}  ($\lambda$, $\delta$, $\mu$, $\phi$, $\gamma$): These
least hard classes exhibit $>54\%$ diskbb contribution with minimal jet
signatures. The nearly Keplerian rotation ($\Omega/\Omega_K \sim 0.95$) and
low magnetization ($\sigma_m \sim 0.1$) favor noticeable cooling and disk
emission over jet launching. The $\lambda$ and $\delta$ classes show steady
accretion with minimal variability beyond stochastic MRI turbulence, while
$\mu$ and $\gamma$ maintain high thermal fractions with weak or absent radio
emission. Our simulations predict $P_{\rm jet} \sim 2 \times 10^{37}$
erg s$^{-1}$ (see Table~\ref{tab:state_properties}) and $L_{\rm X, disk,
max} \sim 1.2 \times 10^{39}$ erg s$^{-1}$ (see
Figure~\ref{fig:spectra_combined}, Table~\ref{tab:state_properties}), with
$L_{\rm X, disk}/P_{\rm jet} \sim 60$. Energy dissipation occurs primarily
through turbulent viscosity in the disk and thermal winds rather than
organized electromagnetic extraction.

It is important to note that typical high/soft states in black hole X-ray
binaries are characterized by geometrically thin, optically thick Keplerian
disks that efficiently radiate thermal emission extending to the innermost
stable circular orbit. However, our advective GRMHD simulations primarily
describe hot, geometrically thick flows that remain optically thin and cannot
produce such soft thermal spectra. In GRS 1915+105, even the ``softer''
classes retain significant hard X-ray components (>40\% power-law
contribution in $\lambda$, $\delta$, $\mu$, $\phi$, $\gamma$ classes),
suggesting that the inner accretion flow remains advective and sub-Keplerian
even during these states. Our model, therefore, explains the persistent hard
emission component arising from the hot inner flow, but does not account for
the outer thin disk thermal emission that produces the soft spectral component.
The SANE state thus represents the regime where the advective inner flow is
weakest with least collimated outflows, allowing maximum contribution from any
outer radiatively efficient disk, rather than a pure soft state transition.
Our ideal GRMHD simulations describe the hot, optically thin advective inner
flow and its hard X-ray emission. The soft thermal component of emission comes
from a geometrically thin outer Keplerian disk, whose radiative cooling cannot
be captured without self-consistent radiative GRMHD treatment
\citep{Chatterjee:2023}. Our model therefore accounts for the hard X-ray
component present in all twelve temporal classes---which constitutes $>40\%$
of the emission even in the softest observed classes \citep{adegoke2018}

Table~\ref{tab:state_properties} summarizes the complete mapping between the
physical properties and the observed classes for GRS~1915+105.

\begin{table}[H]
\small
\caption{Physical properties of MAD, INT, and SANE states for $a = 0.998$,
scaled to GRS 1915+105 ($M = 14\,M_\odot$, $\dot{m} = 0.01$). Spectral
luminosities from Figure~\ref{fig:spectra_combined}.}
\label{tab:state_properties}

\begin{adjustwidth}{-\extralength}{0cm}
\begin{tabularx}{\fulllength}{lCCC}
\toprule
\textbf{Property} & \textbf{MAD} & \textbf{INT} & \textbf{SANE} \\
\midrule
\multicolumn{4}{l}{\textit{Magnetic \& Flow Properties}
} \\
$\phi_{\rm BH}$ & >50 & $20\text{--}40$ & $<20$ \\
$\sigma_m$ (at $r_H$) & 5.89 & 0.48 & 0.13 \\
Plasma-$\beta$ & 0.12 & 1.16 & 3.24 \\
$\Omega/\Omega_K$ & 0.55 & 0.92 & 0.95 \\
$\Gamma_{\rm max}$ & 3.2 & 2.7 & 2.6 \\
\midrule
\multicolumn{4}{l}{\textit{Kinematic Power Budget} 
} \\
$\eta_{\rm jet}$ & $1.0\text{--}1.5$ & $0.3\text{--}0.5$ & $0.1\text{--}0.2$ \\
$P_{\rm jet}$ (erg s$^{-1}$) & $\sim$$2 \times 10^{38}$ & $\sim$$7 \times 10^{37}$ & $\sim$$2 \times 10^{37}$ \\
$L_{\rm X,disk}$ (erg s$^{-1}$) & $\sim$$3.2 \times 10^{39}$ & $\sim$$1.4 \times 10^{39}$ & $\sim$$1.2 \times 10^{39}$ \\
\midrule
\multicolumn{4}{l}{\textit{Radiative Properties (GRS~1915+105)}
} \\
$\nu L_\nu$ (10--100 keV) (erg s$^{-1}$) & $4.8 \times 10^{38}$--$3.2 \times 10^{39}$ & $3.2 \times 10^{38}$--$1.4 \times 10^{39}$ & $1.6 \times 10^{38}$--$1.2 \times 10^{39}$ \\
Peak synchrotron $\nu L_\nu$ (erg s$^{-1}$) & $\sim$$3.3 \times 10^{37}$ & $\sim$$2.4 \times 10^{37}$ & $\sim$$1.1 \times 10^{37}$ \\
Peak inverse-Compton $\nu L_\nu$ (erg s$^{-1}$) & $\sim$$1.4 \times 10^{40}$ & $\sim$$5.3 \times 10^{39}$ & $\sim$$3.8 \times 10^{39}$ \\
\midrule
\multicolumn{4}{l}{\textit{Observational Signatures}
} \\
Variability & High, periodic & Moderate, episodic & Low, stochastic \\
GRS 1915+105 & $\chi$, $\rho$, $\alpha$ & $\nu$, $\kappa$, $\beta$, $\theta$ & $\lambda$, $\delta$, $\mu$, $\phi$, $\gamma$ \\
\bottomrule
\end{tabularx}
\begin{minipage}{\fulllength}
\footnotesize \textit{Note:} Italic row headers denote grouped property categories within the table.
\end{minipage}
\end{adjustwidth}
\end{table}

\subsection{Temporal Variability of X-Ray Emission}
\label{sec:xray_lc}

\begin{figure}[H]
\includegraphics[width=0.95\textwidth]{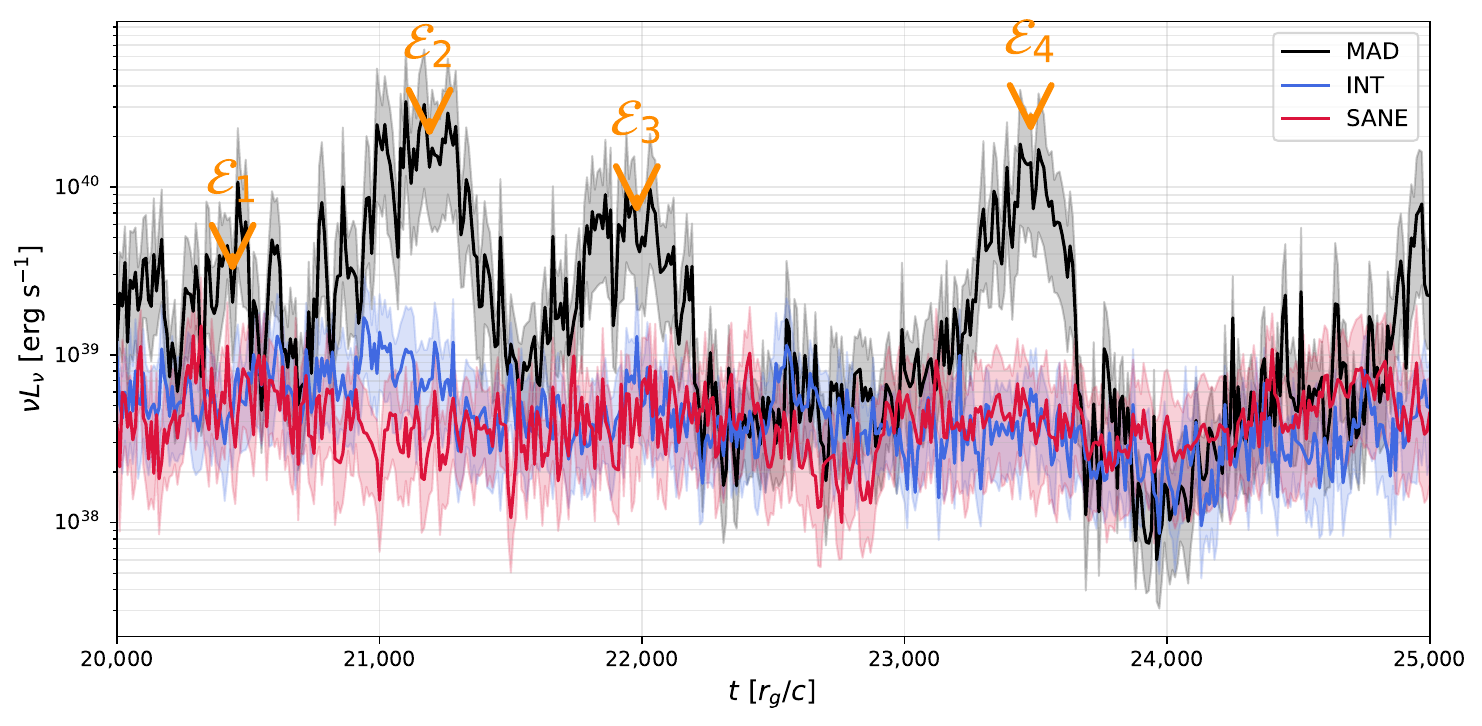}
\caption{Band-integrated
 X-ray luminosity $\nu L_\nu$ in the $10$--$100$~keV
band as a function of simulation time for MAD (black), INT (blue), and SANE
(red) states, scaled to GRS~1915+105 ($M = 14\,M_\odot$, $\dot{m} = 0.01$).
Solid lines show the arithmetic mean over frequency bins within the band;
shaded regions indicate the minimum-to-maximum spread across those bins at
each timestep. Downward arrows mark the four flux eruption events
identified in the MAD light curve (see Section~\ref{sec:xray_lc} for the detection
methodology). The simulation window of $5000\,r_g/c$ corresponds to $\approx
0.35$~s physical time for this source, capturing inner-flow dynamical
timescales rather than the observed class-transition timescales ($\sim
10^5$--$10^6\,r_g/c$).}
\label{fig:xray_lightcurve}
\end{figure}

Figure~\ref{fig:xray_lightcurve} displays the band-integrated X-ray luminosity
$\nu L_\nu$ in the $10$--$100$~keV band as a function of simulation time over
the quasi-steady interval $t = 20{,}000$--$25{,}000\,r_g/c$, computed from
the time-resolved spectral outputs described in Section~\ref{sec:sed_calc}. The
shaded bands indicate the minimum-to-maximum spread across frequency bins
within the band at each timestep, while solid lines show the arithmetic mean.
Table~\ref{tab:xray_lc_stats} summarises the key statistical properties of
each light curve, scaled to GRS~1915+105 ($M = 14\,M_\odot$, $\dot{m} =
0.01$).

\begin{table}[H]
\caption{Statistical properties of the $10$--$100$~keV band-averaged light
curves for MAD, INT, and SANE states, scaled to GRS~1915+105 ($M =
14\,M_\odot$, $\dot{m} = 0.01$). All luminosities are in units of
erg~s$^{-1}$. The fractional variability $\sigma/\mu$ and peak-to-trough
ratio $\nu L_{\nu,\rm max}/\nu L_{\nu,\rm min}$ quantify the amplitude of
temporal fluctuations. GRS~1915+105 class assignments follow the state mapping
of Table~\ref{tab:state_properties}.}
\label{tab:xray_lc_stats}
\begin{tabularx}{\textwidth}{Lcccccc}
\toprule
\textbf{State}
  & \boldmath{$\langle \nu L_\nu \rangle$}
  & \textbf{Median} \boldmath{$\nu L_\nu$}
  & \boldmath{$\nu L_{\nu,\rm min}$}
  & \boldmath{$\nu L_{\nu,\rm max}$}
  & \boldmath{$\sigma/\mu$}
  & \textbf{Max/Min} \\
\midrule
MAD  & $3.33\times10^{39}$ & $1.24\times10^{39}$ & $6.01\times10^{37}$ & $3.23\times10^{40}$ & $1.48$ & $537$  \\
INT  & $4.63\times10^{38}$ & $4.04\times10^{38}$ & $8.60\times10^{37}$ & $1.66\times10^{39}$ & $0.53$ & $19$   \\
SANE & $4.30\times10^{38}$ & $3.86\times10^{38}$ & $1.00\times10^{38}$ & $1.49\times10^{39}$ & $0.45$ & $15$   \\
\bottomrule
\end{tabularx}
\end{table}

The light curves reveal a clear luminosity hierarchy in the hard X-ray band
consistent with the time-averaged spectral energy distributions of
Section~\ref{sec:sed_results}: MAD maintains the highest mean emission $\langle \nu
L_\nu \rangle \approx$ $3.33\times10^{39}$~erg~s$^{-1}$, followed by
INT at $\approx 4.63\times10^{38}$~erg~s$^{-1}$, and SANE at $\approx
4.30\times10^{38}$~erg~s$^{-1}$. As established in Section~\ref{sec:lum_hierarchy},
this ordering is driven by the stronger magnetic fields in MAD configurations
producing a higher synchrotron seed photon field, which in turn elevates the
inverse-Compton output in the 10--100 keV band.

The three states exhibit strikingly different variability characters. The MAD
state is by far the most variable ($\sigma/\mu \approx 1.48$, Max/Min $=
537$), reflecting the dynamically driven nature of accretion under strong
magnetic flux accumulation. The INT and SANE states display comparably moderate
variability ($\sigma/\mu \approx 0.53$ and $0.45$, respectively), both
driven by stochastic MRI turbulence (SANE predominantly so) in their weakly to
moderately magnetized disks, consistent with the irregular, non-periodic
character of the $\gamma$ and $\delta$ classes noted by \citet{adegoke2018}.

A key distinction within the MAD state is the pronounced asymmetry between its
mean and median luminosities: $\langle \nu L_\nu \rangle \approx
3.33\times10^{39}$~erg~s$^{-1}$ versus median $\approx
1.24\times10^{39}$~erg~s$^{-1}$. This skewness demonstrates that the MAD
light curve is dominated by quiescent emission at luminosities $\lesssim
1.5\times10^{39}$~erg~s$^{-1}$, punctuated by rare, short-lived flares that
elevate the mean above the median. We identify four distinct flux eruption
events (shown in Figure~\ref{fig:xray_lightcurve}) using a peak-detection
algorithm applied to the Gaussian-smoothed (smoothing scale $\sigma_{\rm sm} =
5$ timesteps) logarithm of the band-mean light curve, requiring each peak to
rise at least $0.3$~dex above its local baseline. This approach suppresses
local stochastic fluctuations not associated with genuine eruption events. The
four events are distributed across different times indicated by arrows in Figure~\ref{fig:xray_lightcurve} at $t \approx
20{,}440$, $21{,}190$, $21{,}980$, and $23{,}480\,r_g/c$, with X-ray
brightening FWHM durations of $\approx 171$, $438$, $390$, and $392\,r_g/c$
and individual peak luminosities of $2.1\times10^{39}$, $1.3\times10^{40}$,
$4.6\times10^{39}$, and $1.4\times10^{40}$~erg~s$^{-1}$, respectively. These
eruptions correspond to quasi-periodic magnetic flux ejection cycles that are
well established in the MAD literature \citep{tchekhovskoy2011,
Chatterjee:2022}: when the accumulated magnetic flux at the horizon exceeds the
MAD saturation threshold $\phi_{\rm BH} \gtrsim 50$, interchange and
Rayleigh--Taylor-type instabilities expel large flux bundles into the disk,
temporarily reducing $\phi_{\rm BH}$ before renewed accumulation reinstates
the arrested state. The observed quiescent X-ray luminosity of GRS~1915+105 in
the $\chi$ class spans \mbox{$L_X \sim 4$--$15 \times 10^{38}$~erg~s$^{-1}$
\citep{Zdziarski2005, Reid2014}}, consistent with the lower portion of our
simulated MAD range ($\nu L_\nu \sim 4.8 \times 10^{38}$--$3.2 \times
10^{39}$~erg~s$^{-1}$ in the 10--100~keV band). Importantly,
\citet{Punsly2013} demonstrated that major superluminal radio flares in
GRS~1915+105 are systematically preceded by elevated pre-flare X-ray
luminosities reaching $L_X \sim 10^{38}$--$1.5 \times 10^{39}$~erg~s$^{-1}$,
with the jet ejection power strongly correlated with the pre-flare accretion
luminosity. These major flare events---associated with large-scale plasmoid
ejections---map naturally onto our simulated MAD flux eruption events, whose
peak luminosities reach $\sim 10^{40}$--$1.4\times10^{40}$~erg~s$^{-1}$.
This provides a self-consistent physical picture: quiescent $\chi$-class
emission corresponds to the baseline MAD state between eruptions, while the
elevated pre-flare luminosity and subsequent superluminal ejection correspond
to our simulated flux eruption cycle, connecting the kinematic and radiative
signatures of MAD states to observed behaviour across the full dynamic range of
GRS~1915+105 variability.

The rich class phenomenology of GRS~1915+105 is interpreted within our
framework as arising from secular magnetic flux evolution: the gradual
accumulation or dissipation of $\phi_{\rm BH}$ on accretion timescales
provides a natural mechanism for transitions between the MAD, INT, and SANE
states whose intrinsic radiative characters are quantified here. We note that
our simulations fix the accretion state through initial conditions and do not
directly simulate real-time transitions; the transition picture is therefore an
inference from the framework rather than a direct result \citep{raha2026}. The
MAD state achieves its peak hard X-ray luminosity ($\sim
4\times10^{40}$~erg~s$^{-1}$) only during rare magnetically driven
eruptions, while INT and SANE reach substantially lower peaks ($\sim
1.7\times10^{39}$ and $\sim 1.5\times10^{39}$~erg~s$^{-1}$
respectively) sustained by continuous MRI-driven variability. This distinction
between rare-flare-dominated emission in MAD states and lower-level stochastic
emission in INT and SANE states provides an observationally testable
discriminant between accretion states across the GRS~1915+105 variability
classes.

\subsection{X-Ray and Radio Power Correlation}
\label{sec:xray_radio}

\begin{figure}[H]
\includegraphics[width=0.95\textwidth]{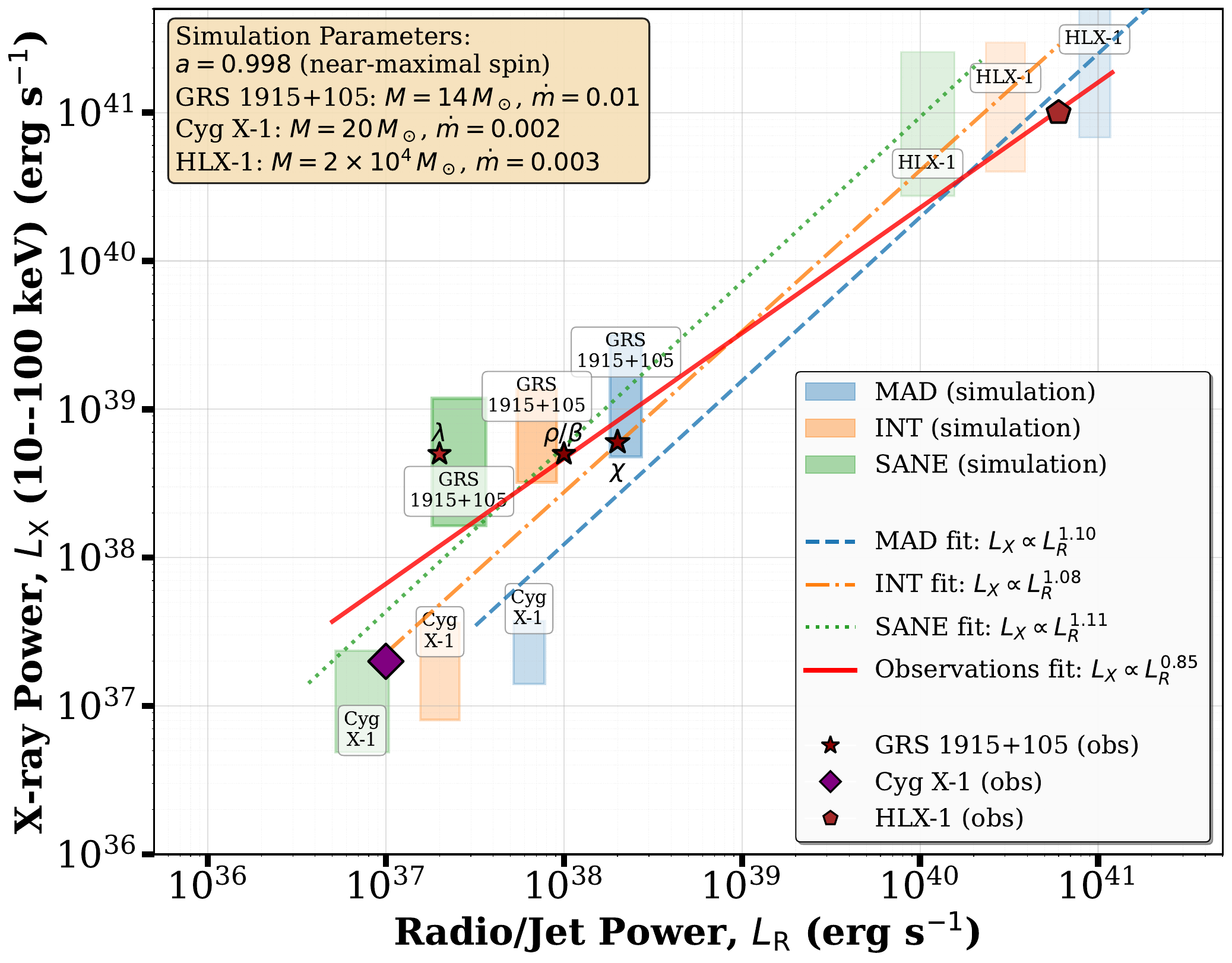}
\caption{X-ray
 (10--100 keV) versus radio/jet power correlation. Colored
regions show simulation predictions for MAD (blue), INT (orange), and SANE
(green) states scaled to GRS~1915+105, Cyg~X-1, and HLX-1. Individual
state fits yield $L_X \propto L_R^{1.10}$ (MAD), $L_{\rm X} \propto
L_R^{1.09}$ (INT), and $L_{\rm X} \propto L_R^{1.11}$ (SANE). The observational aggregate (red line) shows $L_{\rm X} \propto L_R^{0.85}$, consistent with a mixture of states. Observation points: GRS~1915+105 classes
(dark red stars), Cyg~X-1 (purple diamond), HLX-1 (brown\mbox{ pentagon).}}
\label{fig:xray_radio_correlation}
\end{figure}

Here we explore the correlation of X-ray luminosity $L_{\rm X}$ (simulated
$L_{\rm X,disk}$) with radio luminosity $L_R$ (simulated $P_{\rm jet}$).
Figure~\ref{fig:xray_radio_correlation} shows the $L_{\rm X}$--$L_R$ plane
determined by magnetic\mbox{ state.}

Within each magnetic state, both $P_{\rm jet}$ and $L_{\rm X, disk}$ scale
linearly with $\dot{m}$ and $M$, producing near-linear individual correlations
with slopes $L_X \propto L_R^{1.10}$, $L_R^{1.09}$, and $L_R^{1.11}$
for MAD, INT, and SANE, respectively. The three states occupy systematically
different positions in the $L_X$--$L_R$ plane due to their distinct $L_{\rm
X, disk}/P_{\rm jet}$ ratios: SANE is the most X-ray dominated ($L_{\rm
X, disk}/P_{\rm jet} \sim 60$), followed by INT ($L_{\rm X,
disk}/P_{\rm jet} \sim 20$), while MAD shows the smallest ratio ($L_{\rm X,
disk}/P_{\rm jet} \sim 16$) despite having the highest absolute X-ray
luminosity, reflecting MAD's efficient electromagnetic extraction elevating
jet power most strongly among the three states. The observed aggregate slope
$L_X \propto L_R^{0.85}$ emerges from combining observational data
points drawn from sources in different magnetic states rather than from a
single universal mechanism. Sources transitioning between states trace
non-linear paths through the $L_X$--$L_R$ plane.

Our calculated hard X-ray luminosities provide direct observational tests of
this framework. For the MAD classes of GRS~1915+105 ($\chi$, $\rho$,
$\alpha$), our theoretical luminosities $\nu L_\nu \sim 4.8 \times
10^{38}$--$3.2 \times 10^{39}$ erg~s$^{-1}$ in the 10--100 keV band are
consistent with observationally reported values $L_{\rm X} \sim
10^{38}$--$10^{39}$ erg~s$^{-1}$ \citep{Zdziarski2005, Reid2014}. The factor
$\sim 3$ luminosity ratio between MAD and SANE in the hard X-ray band
reflects the stronger magnetic fields in MAD configurations producing higher
inverse-Compton power, as established in \mbox{Section~\ref{sec:lum_hierarchy}.} For the
Cyg~X-1 INT configuration, our predicted \mbox{$\nu L_\nu \sim 8.1 \times
10^{36}$--$3.6 \times 10^{37}$ erg~s$^{-1}$} in the \mbox{10--100 keV} band is
consistent with its observed persistent hard state luminosities \mbox{$\sim
10^{36}$--$10^{37}$ erg~s$^{-1}$ \citep{Grinberg2013}}. HLX-1's extreme hard
X-ray luminosities $\nu L_\nu \sim 6.8 \times 10^{40}$--$5.1 \times
10^{41}$ erg~s$^{-1}$ are explained through efficient Blandford--Znajek
extraction in the MAD configuration combined with its higher black hole mass
\citep{Farrell2009, Godet2012}.

\section{Discussion}\label{sec:discussion}

\subsection{Magnetic Flux as the Fundamental State Variable}

Our simulations demonstrate that the dimensionless magnetic flux $\phi_{\rm BH}$ uniquely determines accretion behavior. Three regimes emerge with well-defined thresholds: MAD ($\phi_{\rm BH} \gtrsim 50$) with magnetic pressure barriers and efficient spin energy extraction; SANE ($\phi_{\rm BH} < 20$) with turbulence-dominated MRI transport; INT ($20 \lesssim \phi_{\rm BH} \lesssim 50$) as a transitional state.

This framework unifies diverse phenomenology: GRS 1915+105's twelve classes arise from rapid magnetic flux evolution; Cyg X-1's steady hard state reflects a stable INT configuration; HLX-1's extreme luminosities result from MAD energy extraction in a massive black hole. The universal applicability stems from dimensionless parameters: $\phi_{\rm BH}$, $\sigma_m$, $\eta_{\rm jet}$; that are mass-independent.

We note again, as mentioned above, that our simulations produce each accretion 
state through distinct initial magnetic field 
configurations and do not directly capture real-time 
transitions between states \citep{raha2026}. The 
transition picture is therefore an inference from 
the framework: advective flux accumulation scaling 
inversely with plasma-$\beta$ provides a plausible 
positive-feedback mechanism driving SANE 
$\rightarrow$ INT $\rightarrow$ MAD evolution on 
accretion timescales $t_{\rm trans} \sim r/v_r$, 
while reverse transitions are consistent with 
violent flux eruptions or gradual Poynting flux 
extraction reducing $\phi_{\rm BH}$ below critical 
thresholds. A direct simulation of state transitions 
requires timescales far exceeding those accessible 
in current GRMHD runs and is identified as a 
primary direction for future work.

\subsection{From Kinematic to Radiative Diagnostics}

 The spectral energy distribution calculations presented in Section~\ref{sec:sed_results} presents the radiative output in the hard X-ray band (10--100 keV), which follows the
MAD $>$ INT $>$ SANE hierarchy. This is consistent with the kinematic energy budgets, with
the physical origin traced to the state-dependent magnetic field strength rather
than electron temperature, as the matched-$T_e$ design of
Figure~\ref{fig:temp_profile} ensures.

The physical connection proceeds through the synchrotron seed photon field.
Stronger large-scale magnetic fields in MAD configurations produce more abundant
high-energy seed photons, driving a correspondingly higher inverse-Compton
power in the hard X-ray band. This dual framework—kinematic
energy diagnostics complemented by radiative calculations with controlled
electron thermodynamics—provides physical insight into energy extraction
mechanisms and quantitative predictions testable against multiwavelength\mbox{
observations.}

\subsection{Observational Validation and Predictions}
\label{sec:obs_validation}

Our predictions match observed luminosities across four orders of magnitude
\mbox{($10^{36}$--$10^{41}$ erg~s$^{-1}$)} when scaled by mass and accretion rate.
The hard X-ray luminosities for GRS~1915+105 MAD states ($\nu L_\nu \sim
4.8 \times 10^{38}$--$3.2 \times 10^{39}$ erg~s$^{-1}$, in 10--100 keV) match
observationally reported hard class values $L_{\rm X} \sim
10^{38}$--$10^{39}$ erg~s$^{-1}$ \citep{Zdziarski2005, Reid2014}, while our
INT configuration for Cyg~X-1 predicts $\nu L_\nu \sim 8.1 \times
10^{36}$--$3.6 \times 10^{37}$ erg~s$^{-1}$, consistent with its persistent
hard state \citep{Grinberg2013}. HLX-1's extreme luminosities are naturally
explained through efficient Blandford--Znajek extraction in the MAD
configuration at higher black hole mass\mbox{ \citep{Farrell2009, Godet2012}.} These
comparisons are discussed in detail in Sections~\ref{sec:lum_hierarchy} and
\ref{sec:grs_classes}.

The observed $L_X$--$L_R$ correlation is reproduced as a superposition of
three magnetic states rather than a single universal mechanism. Each state
individually follows a near-linear correlation ($L_X \propto
L_R^{1.09\text{--}1.11}$) set by their distinct $L_{\rm X,
disk}/P_{\rm jet}$ ratios, while the aggregate of observational data mixing
sources in different states produces the sub-linear slope $L_X \propto
L_R^{0.85}$. The scatter and slope in observed correlations, therefore,
reflect magnetic state diversity across the source population, rather than
accretion rate-dependent efficiency changes. The preserved spectral
characteristics across three orders of magnitude in black hole mass
($14\,M_\odot$ to $2 \times 10^4\,M_\odot$), even via our nonradiative
simulations, demonstrate that the underlying emission physics is governed by
the dimensionless parameters $\phi_{\rm BH}$, $\sigma_m$, and $\Theta_e$,
which are independent of black hole mass.

\subsection{Limitations and Future Work}

Several caveats apply to our framework.

\textbf{Radiative processes:} Our ideal GRMHD approach neglects radiative cooling and radiation pressure. For near- to super-Eddington accretion ($\dot{m} \gtrsim 0.01$), these effects significantly modify flow structure and spectral behaviour. The spectral calculations employ time-averaged simulation data and assumed electron to proton temperature ratios. Radiative GRMHD simulations with self-consistent electron thermodynamics from kinetic plasma physics would eliminate these assumptions \citep{Chatterjee:2023} and enable modeling of spectral variability producing accurate flow spectra on accretion timescales.

\textbf{Spectral Shape:} We cannot predict detailed X-ray spectra without modeling electron heating/cooling. Our spectra given by Figure~\ref{fig:spectra_combined}, particularly for Cyg~X-1, qualitatively corroborate with those in observations \citep{Kantzas2020} and existing model \citep{moscibrodzka2024hardspectralstatexray}. For more quantitative estimate, we need to introduce accurate two temperature evolution of accretion flows and electron distribution. Although, in the present work we consider thermal electron distribution throughout, system can have non-thermal electron distribution, particularly at higher temperature, which would be important to explain the power-law tail of the spectrum. Those rigorous study will be undertaken in future. All these will help reproducing exact frequency and corresponding emission as observed.

\textbf{Simulation timescales:}  Our simulation window of 
$5{,}000\,r_g/c$ corresponds to $\approx 0.35$~s physical time 
for GRS~1915+105 parameters, capturing inner-flow dynamical 
timescales. The observed $\rho$-class ``heartbeat'' oscillation period of 
$\sim$$50$--$100$~s ($\sim$$10^6\,r_g/c$) reflects secular magnetic 
flux accumulation on viscous timescales, which far exceeds the 
duration accessible in our current runs. Our simulated flux 
eruption events capture the correct physical mechanism---MAD 
saturation and interchange instability---but cannot reproduce 
the observed oscillation period, which requires either 
substantially longer simulations or semi-analytic models coupling 
outer-disk flux advection to the inner MAD saturation\mbox{ threshold.}

\textbf{Spin dependence:}  All simulations use 
$a = 0.998$, motivated by observational constraints 
for GRS~1915+105 ($a > 0.98$ \cite{McClintock2006}) 
and Cyg~X-1 ($a \gtrsim 0.98$ 
\cite{Gou2011, Fabian2012}). The qualitative 
MAD/INT/SANE classification and the fundamental 
role of $\phi_{\rm BH}$ are robust to spin: spin 
primarily sets the scale of $\eta_{\rm jet}$ through 
the Blandford--Znajek mechanism \citep{Blandford1977} 
rather than altering the state identity. Lower-spin 
black holes exhibit different MAD saturation 
thresholds and jet efficiencies \citep{Chatterjee:2025}. 
A systematic parameter survey mapping the full 
$(a, \dot{m}, \phi_{\rm BH})$ space 
\citep{Ricarte:2023, Lowell:2024, Chatterjee:2025b} is reserved for 
future work.

\textbf{Binary effects:}  We neglect tidal disturbances, precession, and disk warping that may influence outer disk structure and flux accumulation in close binaries.

Despite these limitations, the core framework—that magnetic flux geometry controls the balance between jet and disk/wind power—appears robust across a wide range of black hole masses and accretion rates. This is particularly true for advective systems, when the accretion rate, and subsequently cooling, hardly influences dynamics. Future work incorporating optical depth effects, radiative transfer, and observer-dependent beaming will enable direct comparison with phase-resolved spectroscopy and polarimetry from IXPE and other next-generation X-ray missions.



\section{Conclusions}\label{sec:conclusions}

We have performed high-resolution three-dimensional GRMHD simulations to
investigate the physical origin of dynamical/temporal and spectral state
diversity in black hole X-ray binaries. Our simulations of a near-maximally
spinning black hole ($a = 0.998$) with varying initial magnetic field
geometries produce three distinct accretion states that explain the rich
phenomenology observed in sources like GRS~1915+105, Cyg~X-1, and HLX-1.

The dimensionless magnetic flux threading the black hole horizon, $\phi_{\rm
BH}$, emerges as the fundamental state variable controlling accretion
behavior. MAD states with $\phi_{\rm BH} \gtrsim 50$ exhibit high
variability, episodic flux eruptions, and efficient jet launching with
\mbox{$\eta_{\rm jet} \sim 1.0$--$1.5$} through Blandford--Znajek extraction of
black hole spin energy. SANE states with $\phi_{\rm BH} < 20$ maintain
steady, low-variability accretion dominated by MRI-driven turbulence with
weak jets ($\eta_{\rm jet} \sim 0.1$--$0.2$). INT states with $20 \lesssim
\phi_{\rm BH} \lesssim 50$ occupy a transitional regime with mixed
properties and inherent instability arising from plasma-$\beta \sim 1$.

The magnetization parameter $\sigma_m$ partitions the flow into three
dynamically distinct regions: magnetically dominated jets ($\sigma_m > 1$)
achieving terminal Lorentz factors $\Gamma \sim 3$--$4$; gas-pressure
dominated disks ($\sigma_m < 1$ within disk scale height) with rotation
profiles ranging from sub-Keplerian (MAD, $\Omega/\Omega_K \sim 0.5$) to
nearly Keplerian (SANE, \mbox{$\Omega/\Omega_K \sim 0.95$)}; and gas-pressure
driven winds ($\sigma_m < 1$ outside the disk region). The $\sigma_m = 1$
surface delineates the jet-wind/disk boundary, with its spatial extent
directly controlled by $\phi_{\rm BH}$.

We validate this kinematic framework through spectral energy distribution
calculations that connect energy fluxes to observable multiwavelength
emission. The proton-to-electron temperature ratios $R = T_p/T_e$ of
150 (MAD), 200 (INT), and 150 (SANE) are chosen to equalize
the electron temperature profiles across states, ensuring that differences
in the emergent spectra reflect the distinct magnetic field structures rather
than differences in electron heating. The resulting inner-disk
dimensionless electron temperatures span $\Theta_e \sim 0.02$--$1.2$
(corresponding to $T_e \sim 10^{8}$--$7\times10^{9}$~K), consistent with
hot, optically thin accretion flows ranging from mildly to highly relativistic
electrons in the innermost disk. The radiative output in the hard X-ray band
(10--100 keV) follows MAD $>$ INT $>$ SANE, with luminosities ranging in
$\nu L_\nu \sim 4.8 \times 10^{38}$--$3.2 \times 10^{39}$ erg~s$^{-1}$
for GRS~1915+105. The factor $\sim 3$ luminosity ratio between MAD and
SANE states reflects the stronger large-scale magnetic fields in MAD
configurations producing a higher synchrotron seed photon energy and
correspondingly stronger inverse-Compton output in the hard X-ray band. The
inverse-Compton spectral peak frequency follows the same MAD $>$ INT $>$ SANE
hierarchy, with peak frequencies $\sim 3.6\times10^{20}$,
$2.7\times10^{20}$, and $2.0\times10^{20}$~Hz, respectively, for
GRS~1915+105, driven by the magnetic field hierarchy setting the
characteristic synchrotron seed photon energies.

Time-resolved X-ray light curves in the 10--100 keV band reveal that this
luminosity hierarchy manifests differently in time: the MAD state maintains
the highest mean hard X-ray emission ($\sigma/\mu \approx 1.48$,
$\langle \nu L_\nu \rangle \approx 3.33\times10^{39}$~erg~s$^{-1}$), while
INT and SANE maintain lower levels ($\approx 4.63\times10^{38}$ and
$\approx 4.30\times10^{38}$~erg~s$^{-1}$, respectively, $\sigma/\mu \approx
0.53$ and $0.45$). The pronounced skewness between MAD's mean
($\approx 3.33\times10^{39}$~erg~s$^{-1}$) and median
($\approx 1.24\times10^{39}$~erg~s$^{-1}$) emission directly reflects its
flare-dominated character. We identify four distinct flux eruption events
using a prominence-based peak detection algorithm applied to the
Gaussian-smoothed logarithm of the band-mean light curve, with X-ray
brightening FWHM durations of $\approx 171$--$438\,r_g/c$ and peak
luminosities of $2.1\times10^{39}$--$1.4\times10^{40}$~erg~s$^{-1}$.

In the $L_X$--$L_R$ plane, SANE states are the most X-ray dominated relative
to their jet power ($L_{\rm X, disk}/P_{\rm jet} \sim 60$), while INT
and MAD show progressively smaller ratios ($L_{\rm X,
disk}/P_{\rm jet} \sim 20$ and $\sim 16$, respectively), reflecting MAD's
efficient electromagnetic extraction elevating jet power most strongly. Each
state individually follows a near-linear correlation ($L_X \propto
L_R^{1.09\text{--}1.11}$), while the aggregate of observational data mixing
sources across states produces the sub-linear slope $L_X \propto
L_R^{0.85}$. The scatter and slope in observed correlations, therefore,
reflect magnetic state diversity across the source population rather than
accretion rate-dependent efficiency changes.

The twelve temporal classes of GRS~1915+105 \citep{belloni2000} map onto our
three magnetic states. MAD produces the hardest classes ($\chi$, $\rho$,
$\alpha$) with strong persistent jets or flares, with the latter quantified
in the time-resolved X-ray light curves (Section~\ref{sec:xray_lc}) as four
distinct flux eruption events with X-ray brightening FWHM durations of
$\approx 171$--$438\,r_g/c$ and peak luminosities reaching $\sim
1.4\times10^{40}$~erg~s$^{-1}$. INT yields transitional classes ($\nu$,
$\kappa$, $\beta$, $\theta$) with rapid spectral variations and episodic jet
activity. SANE includes the least-hard classes ($\lambda$, $\delta$, $\mu$,
$\phi$, $\gamma$) with thermal dominance and weak jets. This unified
framework eliminates the need for separate physical mechanisms for each class,
explaining all observed phenomenology through magnetic flux evolution.
Interestingly, no class of GRS~1915+105 exhibits pure disk-dominated spectra;
at most 54\% diskbb component appears in any class, suggesting that the inner
accretion flow remains advective and sub-Keplerian even during the softest
states. Our advective magnetized simulations, therefore, capture the relevant
inner-flow physics across all twelve classes.

Scaling our simulation results to physical parameters of GRS~1915+105
($M = 14\,M_\odot$, $\dot{m} \sim 0.01$), Cyg~X-1 ($M = 20\,M_\odot$,
$\dot{m} \sim 0.002$), and HLX-1 ($M \sim 2 \times 10^4\,M_\odot$,
$\dot{m} \sim 0.003$) yields predicted powers spanning
$10^{36}$--$\text{few}\times10^{41}$ erg~s$^{-1}$ in excellent agreement
with observations. Both kinematic energy diagnostics and radiative luminosities
scale according to $L \propto \dot{m} M$, with the dimensionless nature of
governing parameters ($\phi_{\rm BH}$, $\sigma_m$, $\eta_{\rm jet}$,
$\Theta_e$) ensuring universal applicability across three orders of magnitude
in black hole mass, even though our simulations are idealized radiation free.

Our results demonstrate that magnetic field geometry---quantified by
$\phi_{\rm BH}$---serves as the fundamental driver of black hole accretion
state diversity. The competition between advective flux accumulation and
dissipative flux removal provides a physical basis for state transitions on
accretion timescales, consistent with observed hysteresis patterns and rapid
variability, though direct simulation of these transitions remains beyond
current computational reach \citep{raha2026}. Future work incorporating
radiation physics, self-consistent electron thermodynamics from kinetic
simulations, and spin-dependent effects will enable detailed spectral
predictions including variability timescales, QPO frequencies, and
polarization signatures testable with IXPE, NuSTAR, and the Event Horizon
Telescope. The spectral energy distributions computed here provide baseline
predictions across radio through hard X-ray bands, while radiative GRMHD
simulations will capture coupling between radiation fields and flow dynamics
essential for super-Eddington accretion. This magnetic flux framework provides
a physical foundation for understanding how black holes accrete, launch jets,
and power some of the most energetic phenomena in the universe.

\vspace{6pt}

\authorcontributions{R.R.~performed the accretion simulations, numerical calculations of spectra, analysis and interpretation, and wrote the first draft. K.C.~performed the accretion simulations, analysis, and edited the draft. 
B.M.~proposed the problem, performed the analysis and interpretation, and thoroughly edited the draft. All authors have read and agreed to the published version of the\mbox{ manuscript.}
}

\funding{R.R.~is supported by the Prime Minister's Research Fellows (PMRF) scheme. K.C. is supported by a CITA postdoctoral fellowship. B.M. would like to acknowledge the project funded by SERB, India, with Ref. No. CRG/2022/003460, for partial support to this research.}

\dataavailability{The data reported in the manuscript are based on rigorous computer simulations and are not publicly available. However, the data could be available from the authors upon reasonable requests.}

\acknowledgments{The authors thank the anonymous referees for providing valuable and profound comments. The authors acknowledge IUCAA Pegasus HPC Cluster for computational resources partially used for this project.}

\conflictsofinterest{The authors declare no conflicts of interest.}
\begin{adjustwidth}{-\extralength}{0cm}

\reftitle{References}

\PublishersNote{}
\end{adjustwidth}
\end{document}